\documentclass[10pt, journal]{IEEEtran}
\usepackage[section]{placeins}
\usepackage{cite}
\usepackage{array}
\usepackage{textcomp}
\usepackage{stfloats}
\usepackage{url}
\usepackage{verbatim}
\usepackage{graphicx}
\usepackage{subfigure}
\usepackage{amssymb}
\usepackage{booktabs}
\usepackage{xcolor}
\usepackage{algorithm}
\usepackage{algpseudocode}
\usepackage{amsthm, amsmath, amsfonts}
\usepackage{bbm}
\usepackage[most]{tcolorbox}
\usepackage{mathrsfs}
\usepackage{multicol}
\usepackage{epstopdf}
\usepackage{listings}

\usepackage{balance}
\lstdefinestyle{ieee-pseudo}{
  basicstyle=\ttfamily\footnotesize,
  breaklines=true,
  columns=fullflexible,
  frame=single,
  numbers=left, numbersep=4pt,
  xleftmargin=0.5em,
  aboveskip=4pt, belowskip=4pt,
  linewidth=\columnwidth
}

\usepackage{enumitem}
\setlist[itemize]{leftmargin=*,itemsep=1.5pt,topsep=2pt,parsep=0pt}
\setlist[description]{leftmargin=1.5em,labelindent=0em,itemsep=2pt,topsep=2pt,parsep=0pt}
\def\BibTeX{{\rm B\kern-.05em{\sc i\kern-.025em b}\kern-.08em
    T\kern-.1667em\lower.7ex\hbox{E}\kern-.125emX}}

\begin{document}
\title{Generative AI Agent Empowered Power Allocation for HAP Propulsion and Communication Systems}
\author{
Xiaoyu Xing,~\IEEEmembership{Graduate Student Member,~IEEE}, 
Dingyi Lu,  
Peng Yang,~\IEEEmembership{Member,~IEEE}, \\ 
Zehui Xiong,~\IEEEmembership{Senior Member,~IEEE}, 
Xianbin Cao,~\IEEEmembership{Senior Member,~IEEE}, and Tony Q. S. Quek,~\IEEEmembership{Fellow,~IEEE}

\thanks{
This study was supported in part by the NSFC under Grant 62471018, in part by the Fundamental Research Funds for the Central Universities, and in part by the National Research Foundation, Singapore and Infocomm Media Development Authority under its Communications and Connectivity Bridging Funding Initiative. 
An earlier version of this paper was presented in part at the 2025 IEEE/CIC International Conference on Communications in China [DOI: 10.1109/ICCC65529.2025.11148893].
The final published version of this article is available in IEEE Transactions on Cognitive Communications and Networking, DOI: 10.1109/TCCN.2026.3683194.
(Corresponding authors: Peng Yang; Xianbin Cao; Tony Q. S. Quek.)

Xiaoyu Xing, Dingyi Lu, and Xianbin Cao are with the School of Electronic Information Engineering, Beihang University, Beijing 100191, China(e-mail: \text{xiaoyuxing; ludingyi; xbcao}@buaa.edu.cn).

Peng Yang is with the School of Electronic Information Engineering, Beihang University, Beijing 100191, China, Pengcheng Laboratory, Shenzhen, 518100, China, and State Key Laboratory of CNS/ATM, Beijing, 100083, China (e-mail: peng\_yang@buaa.edu.cn).

Zehui Xiong is with the School of Electronics, Electrical Engineering and Computer Science, Queen's University Belfast, Belfast BT7 1NN, U.K. (e-mail: z.xiong@qub.ac.uk).

Tony Q. S. Quek is with the Singapore University of Technology and Design, Singapore 487372, and also with the Department of Electronic Engineering, Kyung Hee University, Yongin 17104, South Korea (e-mail: tonyquek@sutd.edu.sg).
}

}

\maketitle

\begin{abstract}
High altitude platforms (HAPs) are emerging as a key enabler for 6G coverage, yet limited energy must be split between propulsion and communications. 
Most prior HAP studies ignore propulsion power or rely on surrogates that miss hull–propeller interference, leading to misestimated communication power budgets and degraded beamforming. 
More importantly, HAP power allocation is intrinsically a multi-system, multidisciplinary problem in which aerodynamics, propulsion-system efficiency, and communication-system performance (quality of service (QoS) and energy efficiency (EE)) are tightly coupled.
To address these challenges, this paper designs an interactive generative artificial intelligence (AI)-empowered HAP power allocation agent.
By interacting with the AI agent, we develop an accurate propulsion power consumption model that takes into account the aerodynamic interference between the HAP's hull and the propeller. 
Assisted by the AI agent, we further formulate a HAP beamforming problem to improve user QoS and enhance the EE of the HAP communication system.
This paper also proposes a QoS-enhanced energy-efficient (Q3E) beamforming algorithm to solve the formulated problem.
Simulation results demonstrate the accuracy of the propulsion-power model and the effectiveness of the Q3E algorithm.
\end{abstract}


\section{Introduction}
\IEEEPARstart{T}{he} rapid increase in demand for seamless connectivity, high capacity, and wider wireless coverage has driven the evolution of wireless communication networks.
Driven by innovations in manufacturing, materials, communications, electronics, and control technologies, high altitude platform (HAP) networks have emerged as essential components of sixth-generation (6G) mobile communication systems \cite{10445467,10745905,cao2024surveynearspaceinformationnetworks,9846953,doi:10.34133/space.0176,AN2025104036}.

Compared with terrestrial networks (TNs), HAPs can respond rapidly to flash-crowd traffic, provide cost-effective coverage for underserved areas, and enhance resilience to natural and man-made disruptions \cite{10680080,10227357}.
A HAP, deployed at an altitude ranging from 17\,km to 25\,km, has the ability to extend communication coverage quickly and provide wide-area communication services.
Compared with satellites, the deployment altitude of a HAP is much lower, resulting in lower communication latency and higher data rates.
Compared with unmanned aerial vehicles (UAVs), a HAP can cover a more expansive area due to its elevated altitude \cite{10673993}. 
However, as emerging networks, many key issues, including channel model and power allocation, must be studied for the HAP networks. 
A HAP incorporates many crucial systems, e.g., propulsion system and communication system. 
This paper is concerned with the propulsion and communication power allocation issue of a HAP.

\subsection{State of the Art}
\textbf{HAP communication power allocation:}
Much research effort \cite{Xu2022RobustMB,10304301,Kong2023UplinkMA,10032267,10330559} has been devoted to the power allocation of the HAP communication system.
For example, the authors in \cite{Xu2022RobustMB} studied the robust beamforming design for intelligent reflecting surface (IRS)-enhanced multiuser downlink communications in a satellite and HAP integrated network, and formulated and solved an optimization problem of minimizing the transmit power of HAP terminals, subject to the constraints on signal-to-interference-plus-noise ratio (SINR) and the outage probabilities experienced at both satellite and HAP terminals.
In \cite{10304301}, the authors proposed a joint user association and beamforming optimization framework for integrated satellite-HAP-ground networks, aiming to maximize the total network throughput while satisfying HAP payload connectivity constraints, HAP and ground base station power constraints, and free-space optical (FSO) backhaul constraints.
In \cite{Kong2023UplinkMA}, the authors proposed to deploy a HAP equipped with a uniform concentric ring array to serve multiple mobile terminals using a space division multiple access (SDMA) technique.
A beamforming scheme aiming at maximizing the minimum average SINR was designed to implement the SDMA technique.
In \cite{10032267}, the authors proposed a deep reinforcement learning (DRL)-based method to enhance the energy efficiency (EE) of HAP networks by optimizing beam vectors, power allocation ratios, and phase shifts, while taking into account the users' quality of service (QoS) requirements and the power limitation of the transmitter.
In \cite{10330559}, the authors investigated the power allocation problem of integrated satellite-HAP-terrestrial networks. 
Taking into account practical limitations like channel estimation errors and imperfect interference cancellation, they proposed an innovative non-orthogonal multiple access-based power allocation strategy to satisfy various QoS requirements of users.

Although the above work \cite{Xu2022RobustMB,10304301,Kong2023UplinkMA,10032267,10330559,10827504,9809985} investigated the optimization issue of communication power consumption of HAP networks, the modeling of propulsion power consumption of a HAP was not touched upon.
Most of them assumed that a HAP did not consume propulsion power when remaining stationary in the stratosphere. 
However, the reality is that to maintain a stationary status in the face of stratospheric winds, the HAP must maintain continuous motion. 
Furthermore, the propulsion system of a HAP consumes a significant amount of power. 
In this case, the modeling of the propulsion power consumption cannot be ignored in the research.
Since a HAP mainly relies on rechargeable batteries and solar panels for its energy supply, unlike TNs, power allocation becomes a critical system design objective.
Accurate propulsion power models are essential for effective power allocation.
Thus, it is essential to investigate the modeling of propulsion power consumption for HAP.

\textbf{HAP propulsion power consumption modeling:}
The modeling of propulsion power consumption has also attracted extensive attention from the research community \cite{ZHU2021106922,Zhang2021MultidisciplinaryDO,Dongchen2023HighaltitudeAP,MOUROUSIAS2024109407,MOUROUSIAS2023108108,SONG2024109266}.
For instance, in \cite{ZHU2021106922}, a propulsion power consumption strategy based on the output power of a solar array and the ambient wind field conditions was proposed.
However, the utilized propulsion power consumption model was not established based on the aerodynamic theory and was not accurate enough.
To improve the accuracy of models, some researchers proposed to derive models of the propulsion power consumption of a HAP based on the aerodynamic theory \cite{Zhang2021MultidisciplinaryDO,Dongchen2023HighaltitudeAP,MOUROUSIAS2024109407,MOUROUSIAS2023108108,SONG2024109266}. 
For instance, the authors in \cite{Zhang2021MultidisciplinaryDO} studied the effects of aerodynamic drag on the propulsion power consumption of a HAP.
In \cite{Dongchen2023HighaltitudeAP}, the authors established a proxy model for calculating a propeller's power consumption and carried out a wind tunnel experiment to verify its correctness.
In \cite{MOUROUSIAS2024109407}, the effects of different kinetic and structural design parameters of HAP's propellers on a propeller's efficiency of a HAP were discussed in detail.
Besides, in \cite{MOUROUSIAS2023108108}, the authors studied the application of polynomial chaos expansion (PCE) method to quantify the uncertainty of the aerodynamic performance of high-altitude propellers.

\subsection{Motivations and Contributions}
Although propeller aerodynamics have been extensively investigated in prior studies \cite{Zhang2021MultidisciplinaryDO,Dongchen2023HighaltitudeAP,MOUROUSIAS2024109407,MOUROUSIAS2023108108,SONG2024109266}, the hull-induced turbulence effects on power consumption of the HAP remain largely unaddressed.
This oversight leads to underestimated propulsion-power requirements, thereby compelling HAP controllers to allocate excessive reserve power to the propulsion system.
These safeguards progressively reduce the power allocated to communication payloads, ultimately constraining the power available for beamforming refinement. 
More importantly, HAP power allocation is intrinsically a multi-system, multidisciplinary problem in which aerodynamics, propulsion-system efficiency, and communication-system performance (QoS and EE) are tightly coupled.
In contrast to existing work, this paper focuses on HAP power allocation through the development of an accurate propulsion-power model and the optimization of communication power consumption. 
We use computational fluid dynamics (CFD)-validated models to compute the propulsion-power budget.
We then pass the resulting feasible communication power to the beamforming optimizer.

The interdisciplinary nature of this research presents significant challenges in formulating mathematical optimization models that accurately represent complex real-world operational scenarios—a task requiring advanced comprehension of nonlinear equations and multi-domain system dynamics.
To address these challenges, we propose a generative AI agent with advanced reasoning capabilities, leveraging techniques including chain-of-thought (CoT) processing and hierarchical problem decomposition \cite{zhou2023leasttomostpromptingenablescomplex}.
Recent applications demonstrate the efficacy of such an AI agent in resolving complex scientific problems, including human genome sequence decoding \cite{GAO20246125}, advanced network security \cite{wang2023makingnetworkconfigurationhuman}, task coordination and scheduling \cite{10839354}, and intelligent system management \cite{CAO2024100570}, aerial access networks \cite{10827504}, UAV networks mobility-aware \cite{9809985}. 
Our main contribution is a generative-AI framework for HAPs that support cross-disciplinary studies.
The key innovations of our work can be summarized as follows:
\begin{itemize}[]
    \item\textbf{Generative AI agent framework design:} 
    This paper proposes a generative AI agent framework for power allocation of a HAP.
    The framework employs a retrieval-augmented generation (RAG) and large language model (LLM) collaboration mechanism.
    This mechanism utilizes a knowledge database constructed by processing academic documents through semantic analysis, text chunking, and embedding techniques.
    It enhances reasoning accuracy through semantic matching between researchers' queries and knowledge database content.

    \item \textbf{Accurate propulsion power consumption modeling:}
    Under the guidance of the generative AI agent, we derive an accurate model for the propulsion power consumption of HAP, utilizing aerodynamic principles and numerical analysis methods.
    More specifically, using CFD numerical analysis techniques for precise aerodynamic assessments, the impact of aerodynamic interference from the HAP's hull on the propulsion system's efficiency is thoroughly analyzed.
    Incorporating these detailed analyses and principles of aerodynamics, an accurate propulsion power consumption model is constructed, which leads to more reliable and accurate communication power constraints. 

    \item \textbf{Optimization of HAP communication power consumption:}
    By interacting with the generative AI agent, we propose a communication power consumption framework for the HAP.
    The framework comprises two key components: 
    1) A HAP beamforming optimization problem that aims to enhance the QoS of terrestrial users and improve the EE of the HAP communication system. 
    2) A QoS-enhanced energy-efficient (Q3E) beamforming algorithm based on a novel artificial neural network (ANN) architecture is designed to solve the formulated problem. 
    By systematically integrating the crucial constraints into the training process, the designed ANN architecture enables unsupervised learning without reliance on labeled datasets. 

    \item Finally, we conduct extensive simulations to verify the accuracy of the propulsion power consumption model and the effectiveness of the proposed beamforming algorithm. 
    Simulation results demonstrate an average 84.3\% reduction in propulsion power consumption deviation for the derived model compared to the existing model.
    The proposed beamforming algorithm demonstrates substantial improvements over the benchmarks, achieving a maximum 33.3\% increase in QoS satisfaction ratio of users and 45\% higher EE. 

\end{itemize}

$Notation$: The upper and lower case boldface letters represent the matrices and column vectors, respectively. 
The definition of the imaginary unit is given by $j = \sqrt{-1}$. 
The representation of unitary space with $M \times N$-dimension is shown as $\mathbb{C}^{M \times N}$. 
The right hand side of $\overset{\Delta}{=}$ denotes the definition of the left hand side. 
The operator of the Kronecker product of two matrices is represented by $\otimes$. 
The symbols $\exp\{ \cdot \}$, $\log_2\{ \cdot \}$, $\ln(\cdot)$ and $\mathbb{E}\{ \cdot \}$ stand for the exponential, base-2 logarithm, natural logarithm and expectation operations, respectively. 
The representations of operating the inverse and Hermitian conjugate are presented as $( \cdot )^{-1}$ and $( \cdot )^H$, respectively. 
The maximum value of a set is denoted by $\max\{ \cdot \}$.
The operator $|\cdot|$ denotes the absolute value for scalars and the cardinality for sets. 
The $l_2$-norm is denoted by $\| \cdot \|_2$.

\begin{table*}[!t]
\centering
\caption{Key Symbols}
\renewcommand{\arraystretch}{0.8}
\small
\begin{tabular}{l p{0.35\textwidth} | l p{0.35\textwidth}}
\toprule
\textbf{{Symbol}} & \textbf{{Meaning}} &
\textbf{{Symbol}} & \textbf{{Meaning}} \\
\midrule
{$P_{\text{HAP}}$} & {Total HAP power} &

 {$\kappa_k$} & {Rician factor}\\
{$P_{\text{prop}}$} & {Propulsion power} &
{$\text{SINR}_k$} & {SINR of user $k$} \\
{$P_{\text{com}}$} & {Communication power} &
{$R_k$} & {Rate of user $k$} \\ 
{$P_{\text{payload}}$} & {Payload power} &
{$EE$} & {Energy efficiency} \\
{$P_{\text{standby}}$} & {Thermal/env.\ \& housekeeping} &
{$\mathcal K,\mathcal Q$} & {User set, QoS-satisfied subset} \\
{$P_{\text{t}}$} & {Static Radio Frequency (RF)/baseband power} &
{$p_k,p_{k,\min}$} & {Power coeff., minimal coeff.} \\
{$\xi$} & {Power Amplifier (PA) inefficiency factor} &
{$P_{\text{tot}}$} & {RF transmit-power budget} \\
{$\mathbf b_k$} & {Beamformer for user $k$} &
{$g_k,\gamma_k$} & {Small-scale gain, avg.\ power}\\
{$\mathbf W$} & {zero-forcing (ZF) beamformer} &
{$T$} & {Aerodynamic drag} \\
{$\mathbf v_k$} & {uniform planar array (UPA) response (user $k$)} &
{$\eta,\eta_p,\eta_m$} & {System/propeller/motor efficiency} \\
{$N_t$} & {Antennas} &
{$C_D,C_{DV}$} & {Total drag coeff., hull drag coeff.} \\
{$K$} & {users} &
{$K_F$} & {Tail drag correction} \\
{$N^x,N^y$} & {UPA elements (x/y)} &
{$\Omega$} & {Hull volume} \\
{$r^x,r^y$} & {Element spacings} &
{$\rho,\mu$} & {Air density, viscosity} \\
{$f_c$} & {Carrier frequency} &
{$l,d,\epsilon$} & {Hull length, width, slenderness} \\
{$B_w$} & {System bandwidth} & 
{$V_0$} & {Airspeed}  \\
{$N_0$} & {Noise power} &  {$\mathrm{Re}$} & {Reynolds number} \\
\bottomrule
\end{tabular}
\end{table*}

\section{Generative AI Agent Framework and HAP Power Consumption Model}

\subsection{Generative AI Agent Framework}

Generative AI agents surpass traditional data interaction models by engaging human users in dynamic, bidirectional information exchange \cite{10531073}.
{In this paper, the agent serves as a retrieval-augmented and validator-gated research assistant.
It compiles and presents literature-based system models and parameter tables to support our HAP study.
The optimization problem is solved by our dedicated solver.}

\begin{figure}[!t]
	\centering
	\includegraphics[width=\linewidth]{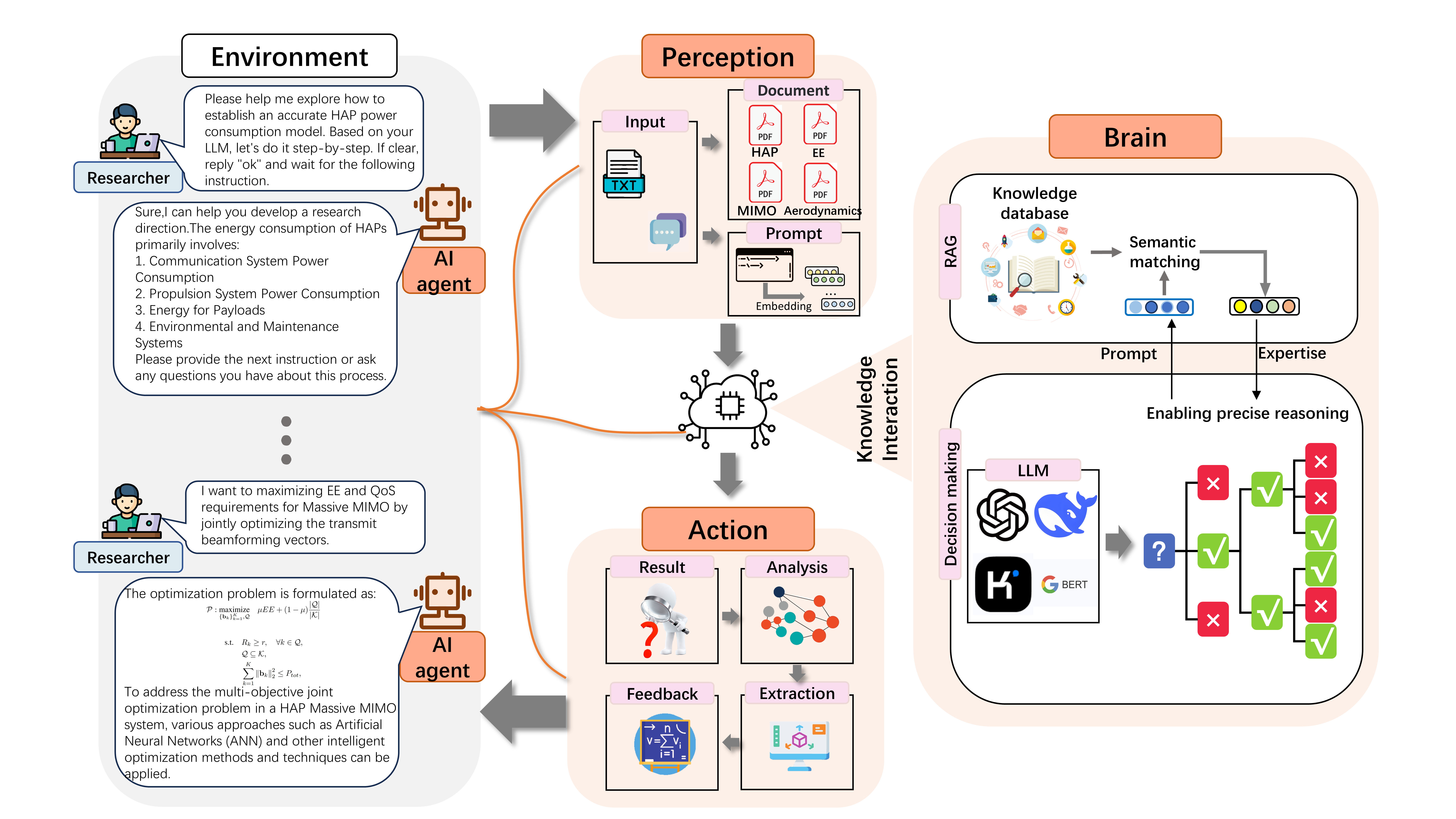}
	\caption{The conceptual architecture of LLM-based generative AI Agent: perception, brain, and action.}
	\label{fig:IAI agent}
\end{figure}

The LLM-based AI agent exemplifies generative AI applications \cite{xi2023risepotentiallargelanguage}, Fig.~\ref{fig:IAI agent} depicts a conceptual structure of an LLM-based generative AI agent, which comprises three components:
1) Perception: This module collects raw input data and converts the data into structured, machine-interpretable semantic representations.
2) Brain: Functioning as the cognitive architecture's central module, it processes information, executes decision-making, and resolves novel tasks through reasoning and planning.
3) Action: Guided by brain-derived inferences, the action module interacts with external environments, analyzes interaction data, and generates contextually appropriate user feedback.
{Below we specify how each module is instantiated in our HAP study.}

\textbf{Perception Module (Data Interpretation):} 
Task-specific prompts are designed to align LLM with interdisciplinary research objectives. 
Raw inputs (e.g., academic documents on massive multiple input multiple output (MIMO), HAP systems, and aerodynamics) undergo semantic analysis, chunking, and embedding. 
This process converts heterogeneous data into structured representations for downstream processing.
{Specifically, PDF documents and notes are divided by a recursive text splitter (chunk size $L_c{=}800$, overlap $L_o{=}150$ tokens), embedded using a normalized sentence encoder, and stored in a Facebook AI Similarity Search (FAISS) index with cosine similarity.} 
{We perform maximal marginal relevance (MMR) selection of $k{=}8$ passages (with $\lambda{=}0.5$) to balance relevance and diversity.
The retrieved context is then concatenated with schema-specific prompts in the brain module.}

\textbf{Brain Module (Knowledge-Augmented Reasoning):}
A knowledge database is constructed from preprocessed textual content, integrating domain theories and empirical constraints. 
The brain module leverages RAG to retrieve contextually relevant information, enabling adaptive reasoning for decision-making. 
This phase ensures academic rigor in deriving theoretical frameworks and methodologies.
In this paper, the brain first extracts the overall HAP system model from the literature. 
It then compiles the parameters and models required by the study according to the research objectives and needs. 
We call the DeepSeek-R1 model through an API, sending the retrieved text chunks and research objectives to the LLM, and use its reasoning ability to support our study.
The main components include:
\begin{itemize}

\item \textbf{CFD atmosphere/flow for setup:} see Sec.~III. 
International Standard Atmosphere (ISA-1976) \cite{9990189} properties at the target altitude, steady incompressible RANS with shear-stress transport (SST) $k$–$\omega$, propellers modeled by Multiple Reference Frame (MRF), and wall resolution $y^+\in[1,5]$.

\item \textbf{Communication model and optimization directions:} A model that links RF transmit power with static terms. 
The solver follows a lexicographic-style solver policy: first maximize the number of QoS-satisfied users $|\mathcal Q|$, then maximize energy efficiency.
\end{itemize}

{To ensure consistency with system tasks and physical laws, the brain performs validation checks: 
1) Pre-processing check: review document provenance and metadata, task scope (HAP altitude envelope, ISA-1976 variants), and unit systems to prevent irrelevant or inconsistent content from entering the index.
2) LLM response admission check: verify consistency of units and dimensions with this paper, assess the feasibility of CFD settings under available computational resources, confirm the aerodynamic plausibility of $P_{\mathrm{prop}}(V_0)$, and ensure that the communication model and optimization objectives are consistent with the HAP network architecture. 
3) End-to-end post-study validation: validate the fitted $P_{\mathrm{prop}}(V_0)$ using CFD data, check curve characteristics within $V_0\in[1,25]$ m/s, confirm RF-budget feasibility with the capped-simplex projector, and cross-check the computation and optimization results against ground test data.}

\textbf{Action Module (Closed-Loop Execution):}
Guided by brain-derived insights, the action module formalizes solutions into mathematical optimization problems and executes them offline in simulation. 
The agent iteratively refines assumptions based on offline experimental feedback (CFD and communication simulations), ensuring alignment with constraints. 
In this paper, the LLM assists with {problem scoping and model drafting}, while numerical CFD and optimization are executed by dedicated solvers rather than the LLM.

\textbf{Technical Limitations and Potential Challenges:}
1) On-platform constraints: HAP platforms operate under strict energy budgets and weight limits; on-board compute cannot host large LLMs, while small-footprint models typically do not meet the accuracy required for safety-critical decisions. Backhaul bandwidth and latency also preclude reliable cloud inference during flight. 
Consequently, our agent is used offline to assist research-stage modeling and parameter preparation; numerical CFD and communication optimization are executed by dedicated solvers, not by the LLM. 
2) Engineer-in-the-loop: Because current LLMs may hallucinate, all AI-inferred components are researcher-reviewed and must pass the admission checks before they parameterize simulations or optimizers.
3) Bias and robustness: AI-inferred components can inherit biases from the literature corpus (e.g., publication bias, dominant-design bias, or outdated ISA-1976 variants) and from retrieval itself (MMR selection bias). 

\subsection{Generative AI Agent-Based Modeling of HAP Power Consumption}

\begin{figure}[!t]
	\centering
	\includegraphics[width=\linewidth]{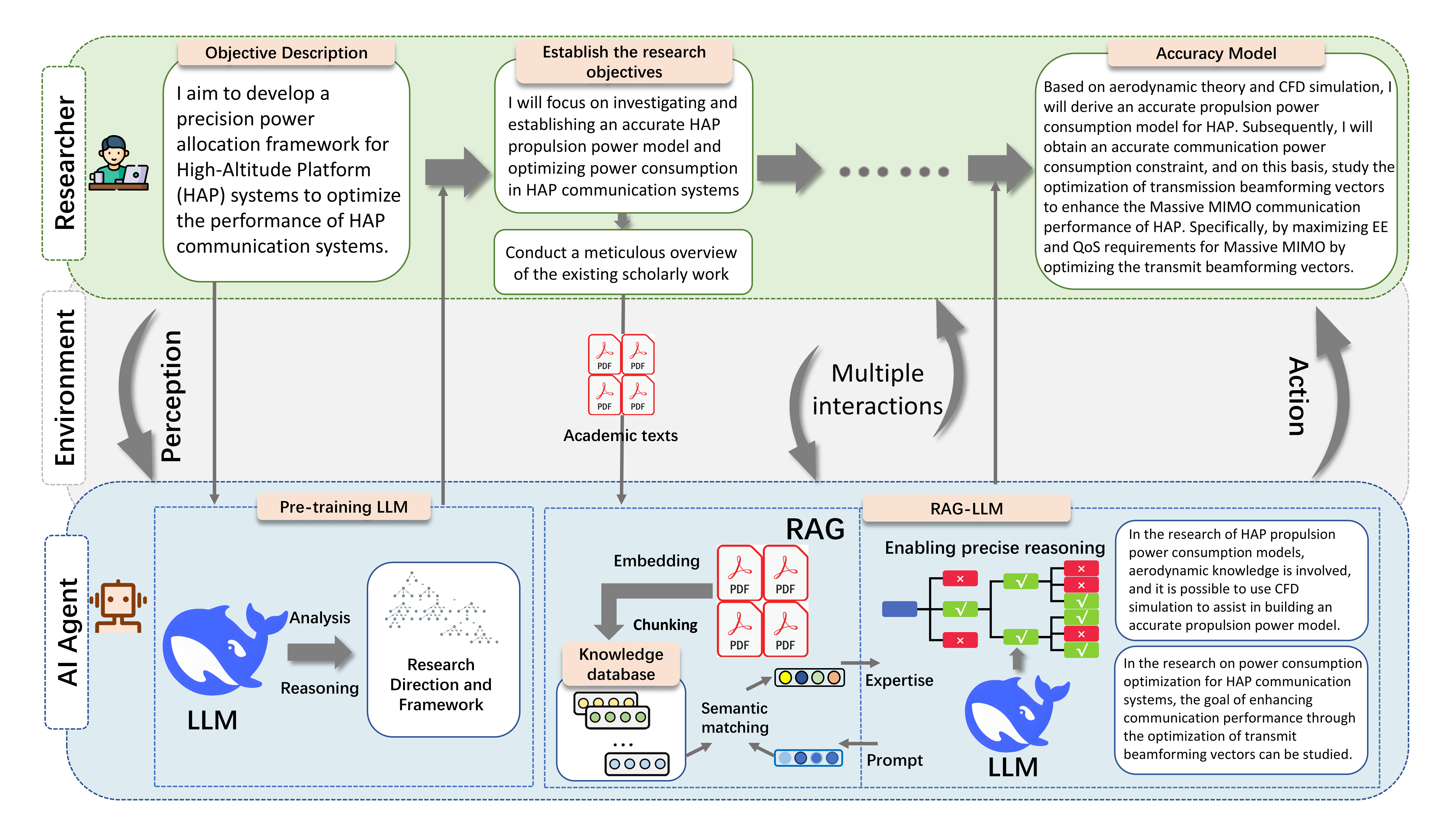}
	\caption{Generative AI Agent-based framework for HAP power consumption modeling.}
	\label{fig:IAI-Based Research}
\end{figure}
{We operationalize the perception–brain–action loop of Sec.~II-A as shown in Fig.~\ref{fig:IAI-Based Research}: 
the perception pipeline prepares the corpus and index, the brain performs RAG-augmented reasoning with validator checks, and the action stage formalizes optimization and runs simulations offline.}

Firstly, we initiate the research by posing questions to a generative AI agent based on the target problems to be investigated. 
The generative AI agent utilizes existing LLM to analyze the existing questions and proposes a research framework.
Specifically, we inquire of the generative AI agent about the HAP system architecture and components of HAP power consumption, and which academic disciplines are involved in studying HAP power consumption.
{Unlike free-form dialogue, we employ structured prompts so that the agent returns literature-cited, parameterized outputs consistent with HAP system characteristics and the propulsion/communication requirements.}

The HAP power consumption model encompasses several key areas: communication system power consumption, propulsion system power consumption, payload system power consumption, and environmental and maintenance system power consumption. 
Mathematically, the power consumption model for HAP is as follows
\begin{equation}\label{eq:hap energy consumption}
P_{\text{HAP}}=P_{\text{com}}+P_{\text{prop}}+P_{\text{payload}}+P_{\text{standby}},
\end{equation}
where $P_{\text{HAP}}$ is HAP power consumption, $P_{\text{com}}$ is HAP communication system power consumption, $P_{\text{prop}}$ is HAP propulsion system power consumption, $P_{\text{payload}}$ is HAP payload power consumption, $P_{\text{standby}}$ is the power consumption of environmental and maintenance systems of the HAP.
Among these, payload power consumption includes devices such as cameras and environmental sensors; the power consumption for environmental and maintenance systems is primarily related to thermal control, and accurate values for these components can be obtained through ground testing\footnote{Due to the stable atmospheric conditions at the operating altitude of HAPs (such as temperature and humidity), the operating conditions of various equipment and thermal control strategies are determined.
Therefore, it is possible to accurately measure the operational power consumption and thermal control power consumption of various devices on board the HAP through ground environmental tests.}.
This paper will focus on exploring the power consumption modeling of the communication system and the propulsion system of the HAP.

Secondly, we refine the prompt words based on the AI agent's research framework and existing research findings, and provide the AI agent with relevant academic documents to enhance the AI agent's understanding and the precision of its reasoning regarding the research content.
Having clarified the research focuses on the power consumption of the communication system and the propulsion system of the HAP, we proceed to review relevant academic research literature, including studies on massive MIMO \cite{9632434,10608095,10757702,9849060,Huang2023QoSAwarePI,9852292}, beamforming \cite{10445467,10387578,10680080,Xu2022RobustMB,10636212,10304301,10757692,10570799}, HAP energy management \cite{10304250,LIU2024103993}, aerodynamics \cite{SONG2024109266,Dongchen2023HighaltitudeAV,DELGADO2024109365, QI2025620,aerospace9010008}, and other pertinent areas.
These scholarly documents are then input into the knowledge database of the LLM to enhance the learning and construction of an accurate AI agent.
{The final parameter tables are versioned only after passing the brain-level conformance checks and are then delivered to the LLM.}

Finally, upon the completion of the knowledge database construction by the AI agent, researchers can interact with the AI agent based on more detailed research objectives and content.
More specifically, in the research of HAP propulsion power consumption, a precise propulsion model can be constructed based on the methods of aerodynamics combined with CFD simulation. 

With the assistance of the AI agent, we can establish the communication power model as follows
\begin{equation}\label{eq:PA_model}
P_{\text{com}}=\xi\sum_{k=1}^{K}\|\mathbf b_k\|_2^2 + P_{\text{t}},
\end{equation}
where $P_{\text{t}}=N_tP_{\text{RFC}}+P_{\text{LO}}+P_{\text{BB}}$.
We consider a fully-digital architecture, so $N_{\mathrm{RF}}=N_t$.
Here, $\xi$ is the PA inefficiency factor, $\mathbf{b}_k$ denotes the transmit beamforming vector for user $k$, and $P_{\text{RFC}}$, $P_{\text{LO}}$, and $P_{\text{BB}}$ are, respectively, the per-chain RF front-end, local oscillator, and baseband circuit powers.

Next, we introduce the modeling of HAP propulsion power consumption and the optimization of HAP communication power consumption in detail in the following two Sections.

\section{Modeling of HAP Propulsion Power Consumption}
{This study considers a finalized HAP configuration whose hull geometry, propeller size, and placement are fixed by prior system-level optimization. 
The learned propeller–efficiency map $\hat{\eta}_p(V_0)$ is therefore design-specific and not intended for cross-geometry generalization. 
All analyses are performed within this design. 
Hull–propeller interference, which isolated-propeller theory cannot capture, is quantified using CFD.
However, the modeling techniques developed in this paper are applicable to the modeling of propulsion power consumption of other HAP configurations.}

An accurate propulsion-power model for a four-propeller HAP is developed using aerodynamic principles and numerical analysis, with a generative-AI agent assisting literature synthesis and parameter selection.

\subsection{HAP Propulsion Power Consumption Model}

For a HAP flying forward at a constant airspeed of $V_0$, its propulsion power consumption can be modeled as follows \cite{SONG2024109266,QI2025620} 
\begin{equation}\label{eq:Existing propulsion energy consumption}
P_{\text{prop}}=\frac{T{{V}_{0}}}{\eta}=\frac{T{{V}_{0}}}{\eta_p\eta_m},
\end{equation} 
where $T$ is the aerodynamic drag of a HAP, ${\eta }$ is the efficiency of a propulsion system, ${\eta_p }$ is the efficiency of propeller, and ${\eta_m }$ is the efficiency of motor.

\begin{figure}[!t]
	\centering
	\includegraphics[width=\linewidth]{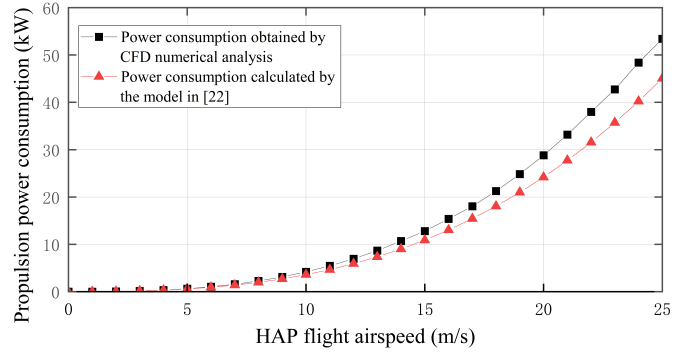}
	\caption{Comparison of power consumption obtained by the model in \cite{SONG2024109266} and numerical analysis.}
	\label{fig:propulsion energy consumption model deviation}
\end{figure}

In order to verify the accuracy of the propulsion power consumption model, the CFD numerical analysis technique is utilized.
Fig.~\ref{fig:propulsion energy consumption model deviation} illustrates the comparison of propulsion power consumption for a HAP at airspeeds $V_0\in[1,25]$ m/s, as obtained from CFD numerical analysis and the model proposed in \cite{SONG2024109266}.
From this figure, we observe a large deviation between the analytical propulsion model and CFD.
Further, the deviation grows superlinearly with airspeed.
For example, when the HAP flight airspeed is 25 m/s, the deviation reaches 8367.3 W.
Therefore, the existing propulsion power consumption model is inaccurate.

The aerodynamic drag depends on environmental parameters, airspeed, and HAP shape.
And the aerodynamic drag model has been extensively studied \cite{SONG2024109266,Dongchen2023HighaltitudeAV,DELGADO2024109365,QI2025620,aerospace9010008}.
Particularly, the aerodynamic drag of a HAP at a constant airspeed of $V_0$ can be calculated by 
\begin{equation}\label{eq:Existing HAP wind resistance}
	T=0.5\rho {{V}_{0}}^{2}{{C}_{D}}{\Omega }^{\frac{2}{3}}=0.5\rho {{V}_{0}}^{2}{{C}_{DV}}{\Omega }^{\frac{2}{3}}{{K}_{F}},
\end{equation}
where $\rho $, ${{C}_{D}}$, ${{C}_{DV}}$, ${{K}_{F}}$, and $\Omega$ denote air density, the HAP drag coefficient, the HAP hull drag coefficient, the tail drag correction factor, and the volume of the HAP.
The HAP hull drag coefficient is given by \cite{HOERNER1965,ALAM2019,MANIKANDAN2021}
\begin{equation}\label{eq:CDV}
	 {{C}_{DV}}=\frac{{0.18{{\epsilon }^{\frac{3}{10}}}+0.27{{\epsilon }^{-\frac{6}{5}}}+1.08{{\epsilon }^{-\frac{27}{10}}}}}{{{{\operatorname{Re}}^{\frac{1}{6}}}}},
\end{equation}
where $\epsilon=l/d $ is the slenderness ratio of a HAP, $l$ the length of HAP, and {$d$} the width of a HAP.
$\operatorname{Re} = \rho {V}_{0} l/\mu$ is the Reynolds number, $\mu$ the {dynamic viscosity}.

\begin{figure}[!t]
	\centering
	\includegraphics[width=\linewidth]{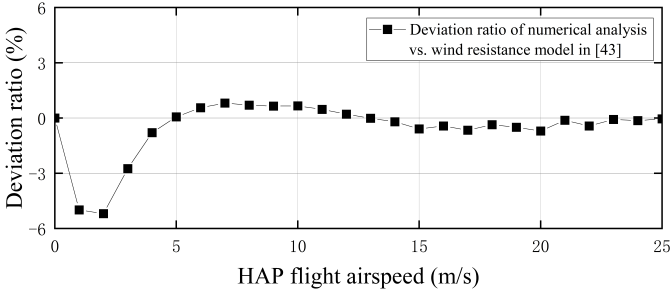}
	\caption{{Deviation ratio of the aerodynamic drag model in \cite{Dongchen2023HighaltitudeAV} versus CFD numerical analysis.}}
	\label{fig:wind resistance deviation ratio}
\end{figure}
\begin{figure}[!t]
	\centering
	\includegraphics[width=\linewidth]{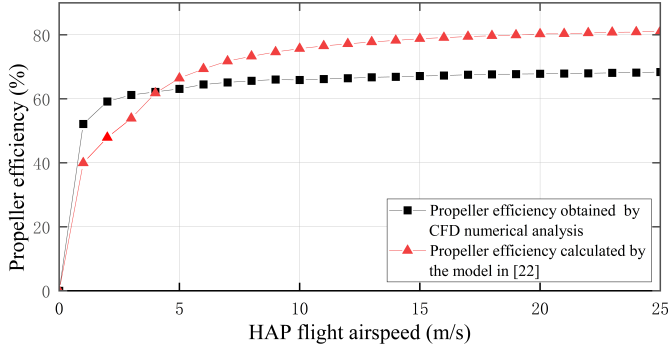}
	\caption{Comparison of propeller efficiency obtained by the model in \cite{SONG2024109266} and numerical analysis.}
	\label{fig:propeller efficiency deviation}
\end{figure}

We conduct a CFD numerical analysis to verify the accuracy of the aerodynamic drag model.
Fig.~\ref{fig:wind resistance deviation ratio} depicts the deviation ratio of the model from the numerical analysis results.
From this figure, we can observe that the achieved deviation ratio of the model is less than 6\%. 
Further, the deviation ratio is within 1\% when the airspeed of the HAP is greater than 5 m/s.

\subsection{Performance Modeling of Propellers}

Similar to (\ref{eq:Existing propulsion energy consumption}), the propeller efficiency of a propeller can be calculated by
\begin{equation}\label{eq:propeller acting thrust_1}
	{{\eta }_{p}}=\frac{{T}_{p}{{V}_{0}}}{{P}_{p}},
\end{equation}
where ${T}_{p}$ is thrust of propeller, ${V}_{0}$ is airspeed of HAP, and ${P}_{p}$ is power consumption of propeller.
To analyze the efficiency of a propeller, we should first analyze the kinetic characteristics of its blade.

{
Consider an $N_b$-blade propeller operating at freestream speed $V_0$ and rotational speed $n_s$. 
Let $r\!\in\![r_0,R]$ denote the spanwise coordinate, $b(r)$ the local chord, $\varphi=\varphi_0+\alpha_i$ the inflow angle, and $(c_l,c_d)$ the sectional lift/drag coefficients. 
Enforcing consistency between blade-element theory and momentum theory yields the axial induction factor \cite{DIEHL202434,Zhong31122024}}
\begin{equation}
a_a=\Bigg(\frac{4K_p\sin^2\!\varphi}{\sigma\,[\,c_l\cos\varphi-c_d\sin\varphi\,]}-1\Bigg)^{-1},\quad
\sigma=\frac{N_b\,b}{2\pi r},
\end{equation}
\begin{equation}
K_p=\frac{2}{\pi}\arccos\!\Big(\exp\!\big\{-\tfrac{N_b(R-r)}{2r\sin\varphi_0}\big\}\Big).
\end{equation}

{With local relative speed $V=V_0(1+a_a)/\sin\varphi$, the thrust and shaft power follow}
\begin{equation}\label{eq:propeller acting thrust}
T_p=\tfrac{1}{2}\rho V_0^2 N_b\!\int_{r_0}^{R}\!\big[c_l\cos\varphi-c_d\sin\varphi\big]\,
b\,\frac{(1+a_a)^2}{\sin^2\!\varphi}\;dr,
\end{equation}
\begin{equation}\label{eq:propeller acting energy consumption}
P_p=\pi n_s\rho V_0^2 N_b\!\int_{r_0}^{R}\!\big[c_l\sin\varphi+c_d\cos\varphi\big]\,
b\,\frac{(1+a_a)^2}{\sin^2\!\varphi}\,r\;dr,
\end{equation}
{and the efficiency is given by \eqref{eq:propeller acting thrust_1}, i.e., $\eta_p=T_pV_0/P_p$. 
These compact expressions are consistent with \cite{DIEHL202434,Zhong31122024} and are used for the subsequent numerical comparison and regression calibration.}

\begin{table*}[t]
\centering
\caption{{CFD setup summary.}}
\label{tab:cfd_setup}
\renewcommand{\arraystretch}{1.0}
\small
\begin{tabular}{p{0.15\linewidth} p{0.8\linewidth}}
\toprule
{Solver/physics} & {Steady incompressible RANS; turbulence: SST $k$-$\omega$; propeller region: MRF} \\
{Domain size} & {$8{\times}5{\times}5$ (streamwise $\times$ spanwise $\times$ vertical) relative to hull length; inlet $2L$ upstream, outlet $5L$ downstream} \\
{BCs} & {Velocity inlet (TI $=1\%$), pressure outlet; no-slip walls} \\
{Mesh} & {Multi-block + BL; total cells $\le 12$M; 12 prism layers; growth $\le 1.2$; target $y^+\in[1,5]$} \\
{Convergence} & {Residuals $\le 10^{-5}$ \& flat thrust/torque/power (drift $<1\%$)} \\
\bottomrule
\end{tabular}
\end{table*}

\begin{figure}[!t]
    \centering
    \subfigure[{Velocity contour of {a propeller operating alone}}]{
        \includegraphics[width=0.8\linewidth]{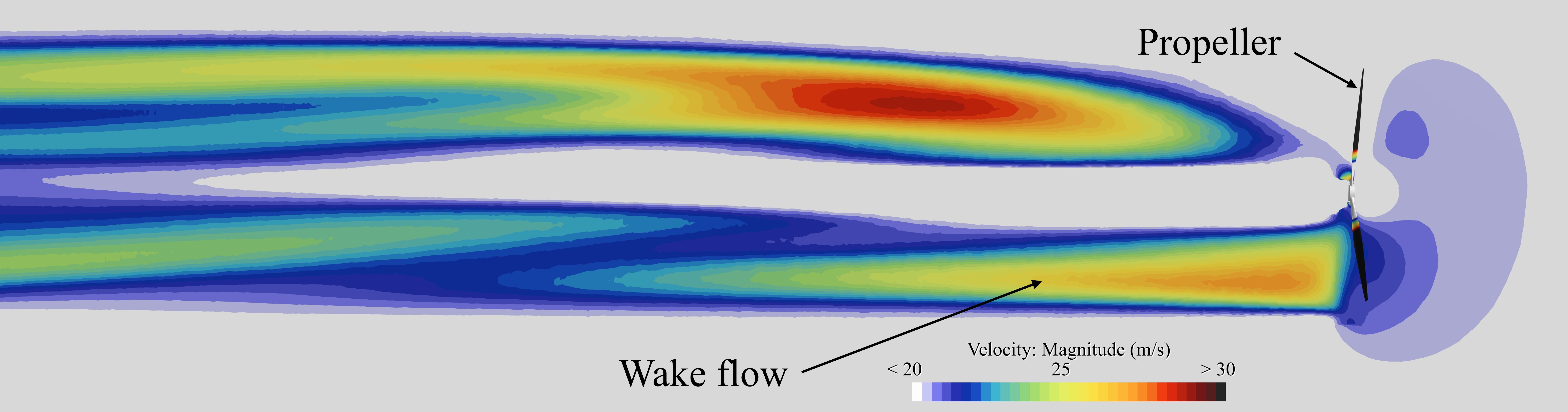}
        \label{fig:bs_subfig:a}}  
    \\  
    \subfigure[{Velocity contour of a propeller installed on a HAP}]{
        \includegraphics[width=0.8\linewidth]{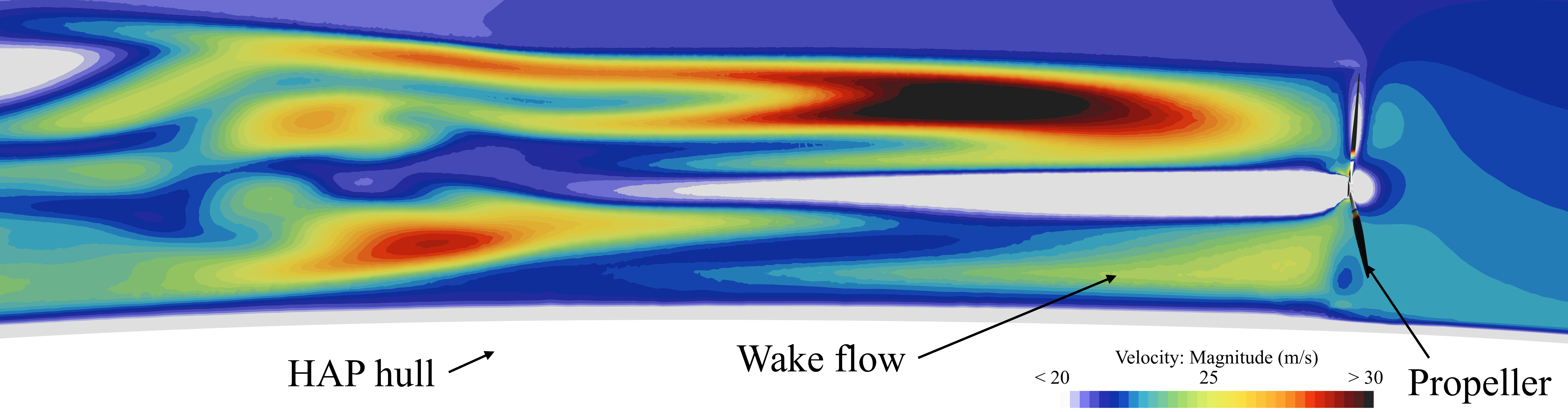}
        \label{fig:bs_subfig:b}
    }
    \\
    \subfigure[{Pressure contour of {a propeller operating alone}}]{
        \includegraphics[width=0.8\linewidth]{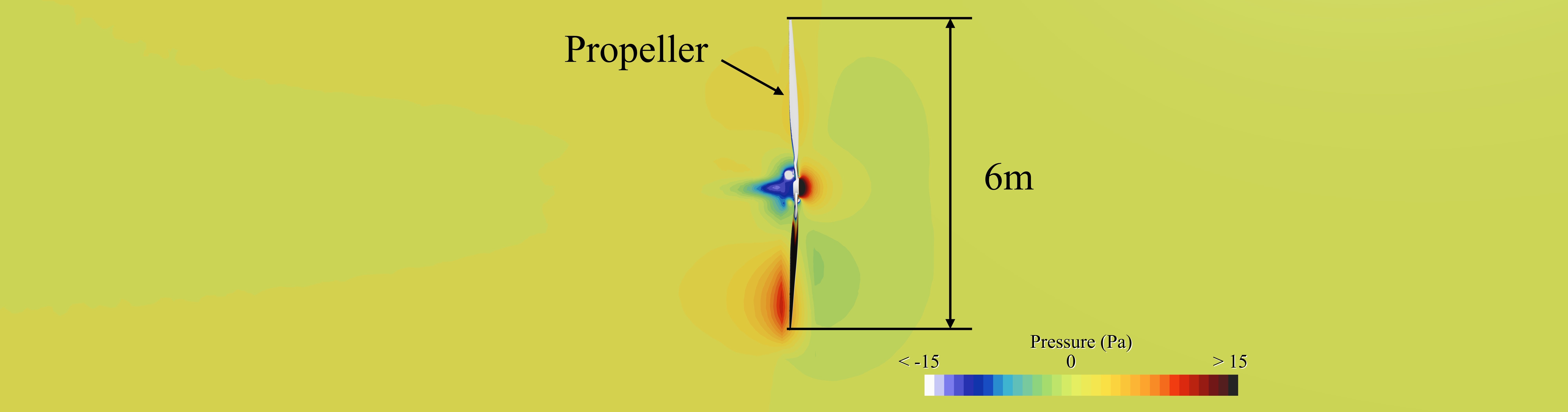}
        \label{fig:bs_subfig:c}}
    \\
    \subfigure[{Pressure contour of {a propeller installed on a HAP}}]{
        \includegraphics[width=0.8\linewidth]{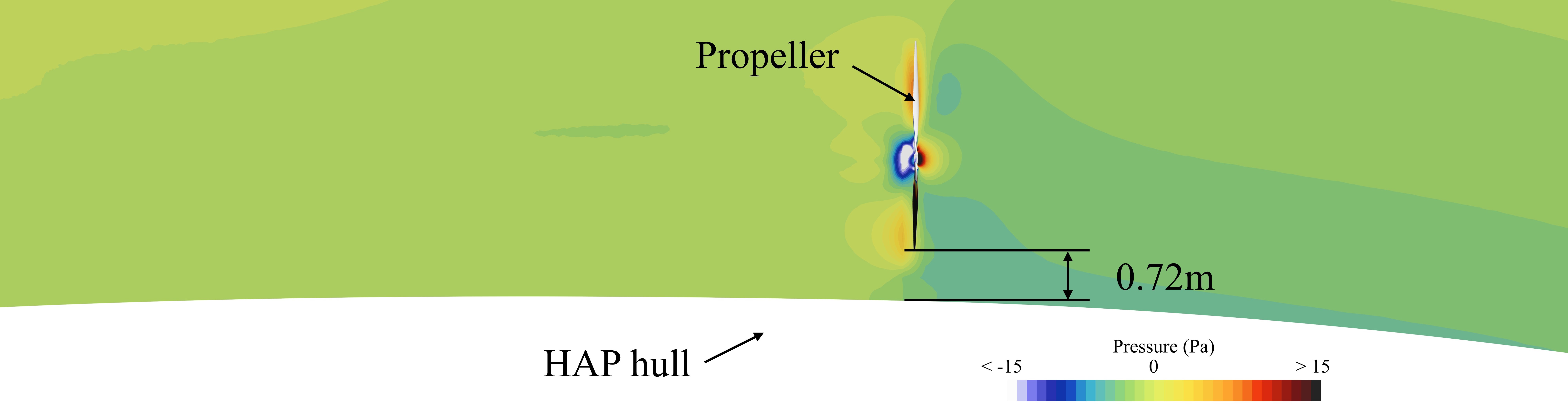}
        \label{fig:bs_subfig:d}}
    
    \caption{Comparison of velocity and pressure contours under different propeller operating conditions at an airspeed of 20 m/s.}
    \label{fig_aerodynamic_interference}
\end{figure}

According to the above theoretical derivations, we can obtain the value of ${\eta }_{p}$.
Next, we verify the accuracy of the derived propeller efficiency model through CFD numerical analysis.
Fig.~\ref{fig:propeller efficiency deviation} illustrates the propeller efficiency obtained by the numerical analysis and the derived model.
The comparison results in Fig.~\ref{fig:propeller efficiency deviation} show that the theoretical model deviates significantly from the numerical analysis. 
For instance, the maximum deviation of the model from the numerical analysis is as high as 12.7\%.
Moreover, the deviation exceeds 10\% when the flight airspeed of a HAP is greater than 12 m/s.
{Because a reliable closed-form proximity correction is not available, we adopt a regression model for $\eta_p(V_0)$ based on CFD analysis.
Over 1-25 m/s, both CFD and the regression indicate a mild increase in $\eta_p(V_0)$.
However, hull proximity lowers its absolute level relative to an isolated propeller.}
We conducted a CFD numerical analysis to explicitly reflect the impact of the HAP hull on its propellers, and the analysis results are shown in Fig.~\ref{fig_aerodynamic_interference}.

From this figure, we can observe that when the propeller operates
alone, the wake flow is regular and the velocity is faster, as shown in Fig.~\ref{fig:bs_subfig:a}.
When a propeller is installed on a HAP, the wake flow is subjected to turbulence, and the wake flow velocity is decreased, as shown in Fig.~\ref{fig:bs_subfig:b}. 
As can be seen from Figs.~\ref{fig:bs_subfig:c} and~\ref{fig:bs_subfig:d}, the
pressure is lower when the propeller is installed on a HAP.
Besides, the closer the propeller is to the HAP hull, the lower the pressure field at the front of the propeller.
From Fig.~\ref{fig_aerodynamic_interference}, we can conclude that the aerodynamic interference caused by the HAP hull is the direct reason affecting the propeller efficiency. 
Analyzing and quantifying the aerodynamic interference is essential to achieving the accurate modeling of HAP propulsion power consumption.

\subsection{Aerodynamic Interference Analysis of HAP Hull}

{As illustrated in Fig.~\ref{fig_aerodynamic_interference}, hull–propeller interference is strongly coupled to geometry and local flow, which makes reliable closed-form corrections difficult. To remain tractable and accurate for the finalized design, we therefore learn the efficiency map $\eta_p(V_0)$ from CFD samples and use this surrogate directly in \eqref{eq:Existing propulsion energy consumption}.}

In this way, we may obtain a tractable expression of the propeller efficiency and the HAP airspeed.
Note that $P_{\text{prop}}(V_0)$ grows superlinearly with airspeed due to the $V_0^3$-like scaling in $T V_0$ and the airspeed-dependent efficiency term.
Therefore, given 25 values of HAP airspeed ranging from 1.0 m/s to 25.0 m/s, we leverage CFD numerical analysis to generate 25 values of the efficiency of the HAP propeller. 
{We adopt an inverse-power surrogate $\hat{\eta}_p(V_0)=c-\alpha V_0^{-\beta}$; on 25 CFD samples it attains low fit errors (e.g., RMSE $=2.33\times10^{-3}$).}
Based on these results, we attempt to capture the non-linear relationship between the propeller efficiency and the airspeed using a non-linear regression technique.
We fit an inverse-power surrogate $\hat{\eta}_p(V_0)=c-\alpha V_0^{-\beta}$ to $n=25$ CFD samples over $V_0\in[1,25]$ m/s, and obtain
\begin{equation}\label{eq:HAP propulsion energy consumption efficiency}
	{\hat \eta}_p=-0.2\,V_{0}^{-\frac{9}{20}}+0.73.
\end{equation}

Finally, by substituting equations (\ref{eq:Existing HAP wind resistance}), (\ref{eq:CDV}), and (\ref{eq:HAP propulsion energy consumption efficiency}) into (\ref{eq:Existing propulsion energy consumption}), we can recalculate the propulsion power consumption of a HAP as
\begin{equation}
\begin{split}
P_{\text{prop}} &
= \frac{T{{V}_{0}}}{{\hat \eta}_p\eta_m}
= 0.5{{\rho }^{\frac56}}{{V}_{0}}^{\frac{17}{6}}{{\mu }^{\frac{1}{6}}}{{\Omega }^{\frac{2}{3}}}\eta_m^{-1}{{K}_{F}} \times\\
\qquad 
&\frac{0.18{{l}^{\frac{2}{15}}}{{d}^{-\frac{3}{10}}}+0.27{{l}^{-\frac{41}{30}}}{{d}^{\frac{6}{5}}}+1.08{{l}^{-\frac{43}{15}}}{{d}^{{\frac{27}{10}}}}}{-0.2{{V}_{0}}^{-\frac{9}{20}}+0.73}.
\end{split}
\label{eq:accurate HAP propulsion energy consumption model}
\end{equation}
{which is used throughout as the core propulsion-power relation for the finalized HAP design.}

Given the obtained propulsion power consumption model, we next investigate the EE of the HAP communication system to achieve the optimization of the HAP communication power consumption.

\subsection{CFD Validation, Uncertainty, and Design-Transfer Workflow}

At a fixed operating altitude (20\,km), we adopt ISA-1976 constant properties $(\rho,\mu)$.
Because the maximum airspeed is $V_0\le 25$~m/s (Mach $<0.1$), validation uses steady, incompressible RANS; turbulence closure and key CFD settings are audited for reproducibility, summarized in Table~\ref{tab:cfd_setup}.

Under identical meshes and boundary conditions in STAR\!-\!CCM+~2502, the propulsion-power differences between SST $k$-$\omega$ and realizable $k$-$\varepsilon$ closures are $1.83\%$ (5\,m/s), $0.90\%$ (15\,m/s), and $0.13\%$ (25\,m/s), i.e., turbulence-model variability $\le 2\%$. Combining this with the validated drag correlation (overall $\le 6\%$, and $\le 1\%$ for $V_0>5$\,m/s) and the surrogate-fit error for $\hat{\eta}_p$ (RMSE $=2.33\times10^{-3}$; relative $\lesssim 0.4\%$ in our range), we obtain the following conservative bound for $V_0>5$\,m/s:
\begin{equation}
\frac{|\Delta P_{\text{prop}}|}{P_{\text{prop}}}\ \lesssim\ 1.0\% + 0.4\% + 2.0\% \ \approx\ 3.4\%.
\label{eq:prop_bound}
\end{equation}

This lever (i.e., 3.4\%) is the uncertainty inherited by the propulsion-informed RF budget in subsequent beamforming optimization.

For a different hull/propeller layout or operating altitude, the efficiency map $\hat{\eta}_p(V_0)$ is re-identified before updating the propulsion-informed RF budget:
\begin{enumerate}
  \item Fix the finalized geometry and propeller placement for the new platform.
  \item Run steady, incompressible RANS CFD over a discrete airspeed grid $V_0 \in \{1,\dots,25\}$\,m/s spanning the admissible envelope at the target altitude (ISA-1976 properties).
  \item Extract $\{\eta_p^{\mathrm{CFD}}(V^{(i)}_0)\}_{i=1}^n$ with convergence checks (residuals $\le 10^{-5}$; flat thrust/torque/power traces).
  \item Fit a smooth, monotone surrogate $\hat{\eta}_p(V_0)$ (e.g., inverse-power family) and validate on holdout speeds.
  \item Update $P_{\mathrm{prop}}=T\,V_0/(\hat{\eta}_p\eta_m)$ and refresh the propulsion-informed RF budget prior to beamforming optimization.
\end{enumerate}

\section{Beamforming Design of HAP Communications}

Guided by a retrieval-augmented LLM agent (Sec.~II), we adopt a lexicographic-style solver policy \cite{HE2021109433} that prioritizes QoS satisfaction over EE: we first enlarge the QoS-satisfied subset under the instantaneous RF budget, and then maximize EE over the resulting feasible set. The diminishing returns of $\log_2(1+\mathrm{SINR})$ motivate prioritizing QoS satisfaction before EE.
According to the results retrieved from the AI agent knowledge database, optimizing power consumption of MIMO systems is crucial for achieving high EE, particularly in scenarios where a large number of antennas and users are involved. 
EE is commonly defined as the ratio of the achievable data rate to the total power consumed \cite{Huang2023QoSAwarePI}. 
A goal of this paper is then to maximize the EE of the HAP communication system and satisfy QoS requirements of users by optimizing the transmit beamforming vectors.
In contrast to directly optimizing a single weighted-sum objective, we follow the same lexicographic-style solver policy motivated by the diminishing-return structure of $\log_2(1+\mathrm{SINR})$.

\subsection{System Modeling of HAP Communications}
We consider a HAP equipped with a UPA consisting of $N^x$ and $N^y$ elements along the x-axis and y-axis, respectively.
Consequently, the total number of antennas is given by $N_t \overset{\Delta}{=} N^x N^y$.
The response vector of the UPA, denoted as $\mathbf{v}_{k}$, can be expressed as
\begin{equation}
\mathbf{v}_{k} = \mathbf{v}_{k}^{x} \otimes \mathbf{v}_{k}^{y} \in \mathbb{C}^{N_{t} \times 1}.
\end{equation}

The array response vector $\mathbf{v}_{k}^{d} \in \mathbb{C}^{N^{d} \times 1}$ for $d \in \mathcal{D} \overset{\Delta}{=}\{ x, y \}$ is defined as \cite{11300920}
\begin{equation}
\begin{split}
\mathbf{v}_{k}^{d} \overset{\Delta}{=} \frac{1}{\sqrt{N^d}} \left [ 1 ,   \exp \big \{-j 2\pi f_{c} \frac{r^{d} u_{k}^{d}}{c} \big \}, \cdots \right.\\
\left. ,\exp \left\{ -j 2\pi f_{c} \frac{r^{d}u_{k}^{d}}{c}(N^{d}-1) \right\} \right ]^{T},
\end{split}
\end{equation} 
where $f_{c}$ represents the carrier frequency, 
and $c$ is the speed of light, 
$r^{d}$ denotes the antenna separation of the UPA, 
$(\theta_{k}^{x}, \theta_{k}^{y})$ represents the angle-of-departure (AoD) of the $k$-th user, {respectively}. 
Then the {spatial-angle} pair of the $k$-th user $(u_{k}^{x}, u_{k}^{y})$ can be represented as 
\begin{equation}
u_{k}^{x}=\sin\theta_{k}^{y}\cos\theta_{k}^{x},
\quad
u_{k}^{y}=\cos\theta_{k}^{y}.
\end{equation}

We assume that $K$ single-antenna users request communication services from the HAP, and the set of users is defined as $ \mathcal{K} \overset{\Delta}{=} \{1, 2, \ldots, K\} $. 
Due to the high altitude of the HAP, the relative positions between the HAP and the 
$k$-th user remain nearly unchanged over a given time period.
In this work, the channel gain $ g_k \in \mathbb{C}$ is modeled to follow the Rician fading distribution random variable with Rician factor $\kappa_k$ and power $\gamma_k$. 
The downlink channel for $k$-th user can be expressed as 
\begin{equation}
\mathbf{h}_{k} = \mathbf{v}_{k} g_{k} \in \mathbb{C}^{N_{t} \times 1}.
\end{equation}

It is worth noting that the channel model adopted in this work is similar to RF channel model described in \cite{10304301}.
In this paper, we utilize statistical channel state information (sCSI), which remains stable over an extended period, for the design of the beamforming scheme.
The received signal for $k$-th user can be expressed as \cite{10636212}
\begin{equation}\label{eq:received signal}
y_k = \mathbf{h}_k^H \sum_{\ell=1}^{K} \mathbf{b}_\ell s_\ell + n_k = \mathbf{h}_k^H \mathbf{b}_k s_k + \mathbf{h}_k^H \sum_{\ell \neq k} \mathbf{b}_\ell s_\ell + n_k,
\end{equation}
where $s_{k}$ denotes the transmitted signal for $k$-th user, $n_{k}$ represents the additive white Gaussian noise (AWGN) with an average energy of $N_{0}$, and $\mathbf{b}_{k}\in \mathbb{C}^{N_{t} \times 1}$ denotes the beamforming vector for $k$-th user. 
Based on (\ref{eq:received signal}), the received SINR of the $k$-th user can be expressed as
\begin{equation}
\text{SINR}_k  \overset{\Delta}{=} \frac{ \left| \mathbf{b}_k^H \mathbf{h}_k \right|^2 }{\sum_{\ell \neq k} \left| \mathbf{b}_\ell^H \mathbf{h}_k \right|^2 + N_0}.
\end{equation}

The beamforming vectors are constrained by the available RF transmit-power budget $P_{\text{tot}}$ \cite{9849060}, which can be expressed as 
\begin{equation}\label{eq:Pbudget}
\sum_{k=1}^{K}\|\mathbf b_k\|_2^2 \le 
\frac{P_{\text{HAP}}-P_{\text{prop}}-P_{\text{payload}}-P_{\text{standby}}-P_{\text{t}}}{\xi}
\ \triangleq\ P_{\text{tot}}.
\end{equation}

\subsection{Problem Formulation}\label{subsec:problem}
In this paper, a beamforming strategy is devised by jointly considering the system's EE and QoS satisfaction ratio.
The user set $\mathcal{K}$ is divided into two subsets:
$\mathcal{Q}$ denoting users with satisfied QoS requirements, and $\mathcal{K} \setminus \mathcal{Q}$ containing the remaining users.  
The QoS satisfaction ratio is mathematically defined as $\frac{|\mathcal{Q}|}{|\mathcal{K}|}$.
Moreover, the system EE is expressed as the ratio between the achievable sum-rate and the total power consumption, formulated as \cite{Huang2023QoSAwarePI}
\begin{equation}
EE = \frac{\sum_{k=1}^{K} R_{k}}{P_\text{com}},
\end{equation}
where $R_{k}$ denotes the average data rate of the $k$-th user, expressed as 
\begin{align}
\begin{split}
R_{k} & 
\overset{\Delta}{=} B_w\,\mathbb{E}\!\left\{\log_2\!\left(1+\text{SINR}_k\right)\right\} \\
&
\overset{\Delta}{=} B_w\,\mathbb{E}\left\{\log_2\left(1 + \frac{\left|\mathbf{b}_{k}^{H}\mathbf{h}_{k}\right|^{2}}{\sum_{\ell \neq k} \left|\mathbf{b}_{\ell}^{H}\mathbf{h}_{k}\right|^{2} + N_{0}}\right)\right\}.
\end{split}
\end{align} 

This paper aims to jointly maximize both the QoS satisfaction ratio and the EE of the HAP communication system.
The optimization problem is formulated as
\setcounter{equation}{22}
\begin{subequations}
\begin{align}
\mathcal{P}: \quad &\underset{\left\{\mathbf{b}_{k}\right\}_{k=1}^{K}, \mathcal{Q}}{\text{maximize}} \quad \omega EE + (1 - \omega) \frac{|\mathcal{Q}|}{|\mathcal{K}|} \label{eq:optimization problem QOSEE}\\
&\text{s.t.} \quad  R_{k} \geq r_{k}, \quad \forall k \in \mathcal{Q}, \label{eq:23a} \\
& \quad\quad\mathcal{Q} \subseteq \mathcal{K},\label{eq:23b}  \\
& \quad\quad\sum_{k=1}^{K} \left\|\mathbf{b}_{k}\right\|_{2}^{2} \leq P_{\text{tot}}, \label{eq:23c}
\end{align}
\end{subequations}
where $r_k$ (in bps) denotes the QoS requirement of user $k$, and $\omega$ is a weight coefficient.
Since $B_w$ is in Hz, $r_k/B_w$ is in bit/s/Hz, consistent with $\log_2(\cdot)$.

\subsection{Beamforming Algorithm Design}

To simplify the expected calculation, we utilize Jensen's inequality to estimate the achievable data rate with an upper bound approximation $R_k$, as follows
\begin{equation}
\begin{split}
R_k \leq \bar{R}_k & \overset{\Delta}{=} B_w\,\log_2 \left( 1 + \frac{\mathbb{E} \left\{ \left| \mathbf{b}_k^H \mathbf{h}_k \right|^2 \right\}}{\mathbb{E} \left\{ \sum_{\ell \neq k} \left| \mathbf{b}_\ell^H \mathbf{h}_k \right|^2 \right\} + N_0} \right) \\
&= B_w\,\log_2 \left( 1 + \frac{\gamma_k \left| \mathbf{v}_k^H \mathbf{b}_k \right|^2}{\sum_{\ell \neq k} \gamma_k \left| \mathbf{v}_k^H \mathbf{b}_\ell \right|^2 + N_0} \right ).
\end{split}
\end{equation}

In the remainder of this section, we use $\bar{R}_k$ as a tractable surrogate of the ergodic rate.
For notational simplicity, we drop the bar and denote the surrogate rate by $R_k$ hereafter.

By employing the ZF method, we effectively mitigate the cross-interference among different users \cite{10192088}, and we define a digital beamformer $\mathbf{W}$, as follows
\begin{equation}
\mathbf{W} = \mathbf{V} \left( \mathbf{V}^H \mathbf{V} \right)^{-1} \in \mathbb{C}^{N_t \times K},
\end{equation}
where $\mathbf{V} = \{\mathbf{v}_{1}, \mathbf{v}_{2}, \ldots, \mathbf{v}_{K}\} $.
Then the beamforming vector $b_{k}$ for user $k$ can be calculated as follows
\begin{equation}
\mathbf{b}_{k} = p_{k}\mathbf{w}_{k},
\end{equation}
where $\mathbf{w}_{k}$ is the $k$-th column of the matrix $\mathbf{W}$, and $p_k$ is the $k$-th user power coefficient.

Thus, we can transform the optimization problem (\ref{eq:optimization problem QOSEE}) into the following
\setcounter{equation}{26}
\begin{subequations}
\begin{align}
\mathcal{P}:\quad &\underset{\left\{p_{k}\right\}_{k=1}^{K}, \mathcal{Q}}{\text{maximize}} \quad \omega \frac{\sum_{k=1}^{K} R_{k}    }{P_\text{com}} + (1 - \omega) \frac{|\mathcal{Q}|}{|\mathcal{K}|}
\label{eq:optimization problem pp}\\
&\text{s.t.} \quad  R_{k} \overset{\Delta}{=} B_w\,\log_2 \left( 1 + \frac{\gamma_k |p_k|^2}{N_0} \right) \geq {r_k}, \quad \forall k \in \mathcal{Q}, \label{eq:27a} \\
& \quad\quad \mathcal{Q} \subseteq \mathcal{K}, \label{eq:27b}  \\
& \quad\quad \sum_{k=1}^{K} \left\| p_{k}\mathbf{w}_{k}\right\|_{2}^{2} \leq P_{\text{tot}}. \label{eq:27c}
\end{align}
\end{subequations}

The logarithmic relationship between rate and SINR ($R_k \propto \log_2(1 + \text{SINR}_k)$) leads to a strictly concave power-rate curve, implying diminishing marginal returns in power efficiency.
Beyond certain SINR thresholds, achieving a unit rate improvement requires superlinear power growth. 
Consequently, we follow a lexicographic-style solver policy: (1) determine the largest QoS-satisfied subset $\mathcal{Q}$ via the feasibility test below; (2) with $\mathcal{Q}$ fixed, maximize $EE$ over the corresponding feasible set under (\ref{eq:27c}).
Under the ZF structure with sCSI, let $p_{k,\min}$ be the minimal per-user powers defined in (\ref{eq:pmink}), and let $C(\mathcal S)\!=\!\sum_{k\in\mathcal S}\|p_{k,\min}\mathbf w_k\|_2^2$ be the minimum budget to make users in $\mathcal S$ feasible. Then selecting users by the ascending order of $\|p_{k,\min}\mathbf w_k\|_2^2$ (greedy) maximizes $|\mathcal Q|$ subject to $C(\mathcal Q)\le P_{\mathrm{tot}}$.
Since $C(\cdot)$ is additive and nonnegative, the problem reduces to a cardinality maximization under a single knapsack with items of weights $\|p_{k,\min}\mathbf w_k\|_2^2$ and identical unit values. The greedy by nondecreasing weights attains an optimal solution. 
With $|\mathcal Q|$ fixed, maximizing EE over the feasible face yields the optimal solution of the second stage under the proposed decomposition.

The minimum required power coefficient $p_{k,\min}$ to satisfy QoS requirements $r_{k}$ for the $k$-th user can be calculated by
\begin{equation}
p_{k,\min} = \sqrt{\frac{N_0 \left(2^{\,r_k/B_w} - 1\right)}{\gamma_k}}.
\label{eq:pmink}
\end{equation}

This follows from $R_k=B_w\log_2(1+\mathrm{SINR}_k)$ under the ZF structure.

Based on (\ref{eq:pmink}), the minimum transmit power required to satisfy the QoS requirements of all users is given by $P_{e_{min}}=\sum_{k=1}^{K} \left\| p_{k,\min}\mathbf{w}_{k}\right\|_{2}^{2}$.
Accordingly, the constraint in (\ref{eq:27c}) can be rewritten as
\begin{equation}
P_{e_{min}} \leq P_{\text{tot}}.
\label{eq:pemin}
\end{equation}

This formulation leads to two distinct cases, as described below:

\textbf{Case 1 (Full User QoS Satisfaction):} 
If (\ref{eq:pemin}) holds, all users achieve or exceed their QoS requirements ($R_{k}\geq r_k, \text{ } \forall k \in \mathcal{K}$), resulting in $|\mathcal{Q}| = |\mathcal{K}|$.
Then problem $\mathcal{P}$ can be transformed into

\setcounter{equation}{29}
\begin{subequations}
\begin{align}
\mathcal{P}_{1}: \quad & \underset{\left\{p_{k}\right\}_{k=1}^{K}}{\text{maximize}} \quad \frac{\sum_{k=1}^{K} R_{k}}{P_{\text{com}}} \label{eq:optimization problem p1} \\
& \text{s.t.} \quad p_{k} \geq p_{k_{\text{min}}}, \quad \forall k \in \mathcal{K}, \label{eq:29a} \\
& \quad \quad\sum_{k=1}^{K} \left\| p_{k} \mathbf{w}_{k} \right\|_{2}^{2} \leq P_{\text{tot}}. \label{eq:29b}
\end{align}
\end{subequations} 

\textbf{Case 2 (Partial QoS Satisfaction):}
If (\ref{eq:pemin}) is violated, at least one user fails to meet its QoS requirement ($\mathcal{Q} \subset \mathcal{K}$).
To maximize the number of QoS-satisfied users $|\mathcal{Q}|$, a prioritization strategy is adopted: 
The magnitude of the power allocated to all users is sorted in ascending order based on their minimum power demands. Let $k=g(i)$, where the function $g: i \to k$, $i\in \mathcal {K}$, and we sequentially allocate power to users $\{g(i)\}$. 
Mathematically, users are ordered according to the following expression
\begin{equation}
\left\| p_{{g(1)}} \mathbf{w}_{{g(1)}} \right\|_2^2 \leq \ldots \le \left\| p_{{g(i)}} \mathbf{w}_{{g(i)}} \right\|_2^2 \le \ldots \leq
\left\| p_{{g(K)}} \mathbf{w}_{{g(K)}} \right\|_2^2.
\label{ascending_order_pk}
\end{equation}

This ordering scheme ensures that users with lower power requirements are prioritized, thereby systematically expanding $|\mathcal{Q}|$.
{The above ordering can be interpreted as a greedy solution to the cardinality-constrained feasibility problem over (\ref{eq:27c}); it is optimal for the ZF-based structure because each user's minimal feasible power contributes additively to (\ref{eq:27c}).}
The maximum admissible set is then determined by
\begin{equation}
\mathcal{Q} = \left\{ {g(1)},...,{g(m)} \right\},
\label{calculate_Q}
\end{equation}
where $ m= \max\left\{ n \Big| \sum_{i=1}^n \left\| p_{{g(i)}} \mathbf{w}_{{g(i)}} \right\|_2^2  \leq P_{\text{tot}} \right\}$.

After allocating minimum required power $p_{k_{\min}}$ to users in $\mathcal{Q}$, the remaining power budget
 $P_{\text{tot}}' = P_{\text{tot}}-\sum_{i=1}^{m} \left\| p_{{g(i)}} \mathbf{w}_{{g(i)}} \right\|_2^2 $
is utilized to enhance EE for users in $\mathcal{K} \setminus \mathcal{Q}$.
Then problem $\mathcal{P}$ can be transformed into
\setcounter{equation}{32}
\begin{subequations}
\begin{align}
\mathcal{P}_2:\quad & \underset{\left\{p_{k}\right\}_{k \in \mathcal{K} \setminus \mathcal{Q}}}{\text{maximize}} \quad \frac{\sum_{k \in \mathcal{K} \setminus \mathcal{Q}} R_{k}}{P_{\text{com}}} \\
&\text{s.t.} \quad \sum_{k \in \mathcal{K} \setminus \mathcal{Q}} \left\| p_k \mathbf{w}_k \right\|_2^2 \leq P_{\text{tot}}'. \label{eq:38a}
\end{align}
\end{subequations}

\begin{figure}[!t]
	\centering
	\includegraphics[width=0.8\linewidth]{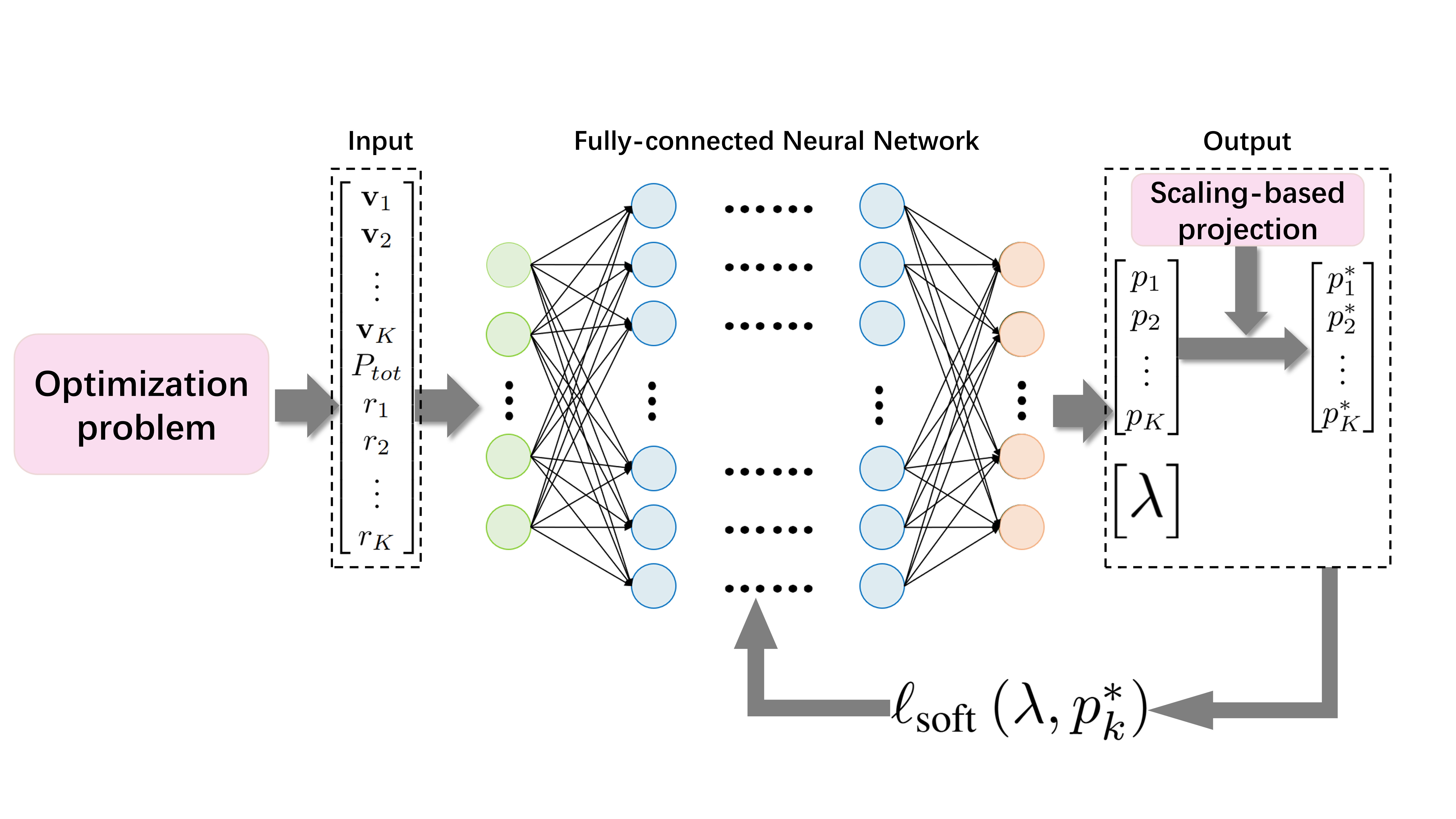}
	\caption{{The ANN architecture that has been contemplated provides a resolution to Problem (\ref{eq:optimization problem pp})}}
 	\label{fig:ANN}
\end{figure}

Both problems $\mathcal{P}_1$ and $\mathcal{P}_2$ are non-convex optimization problems. 
The AI agent generates multiple candidate approaches for their solutions, including the sequential least squares quadratic programming (SLSQP) and ANN methods. 
This paper employs the SLSQP method and develops an ANN method to solve the optimization problem, and compares the performance of these methods. 

\begin{algorithm}
\caption{{Training Loop for the ANN Solver}}
\begin{algorithmic}[1]
\Require {sCSI $\{\gamma_k,\mathbf V\}$, QoS $\{r_k\}$, budget $P_{\mathrm{tot}}$, Adam, max\_epochs$=2000$, patience $P{=}50$.}
\State {Precompute: $\mathbf W = \mathbf V(\mathbf V^H\mathbf V)^{-1}$,\; $p_{k,\min}=\sqrt{\frac{N_0(2^{r_k/B_w}-1)}{\gamma_k}}$.}
\Function{{Net}}{$\mathbf V,\{\gamma_k\},\{r_k\},P_{\mathrm{tot}}$}
  \State {\Return raw $\tilde{\mathbf p}\in\mathbb{R}^K$ from the feed-forward MLP (Sec.~IV-B).}
\EndFunction
\Function{{Project}}{$\tilde{\mathbf p},\mathcal Q$}
  \State $m_k = p_{k,\min}$ if $k\!\in\!\mathcal Q$ else $0$.
  \State $\hat p_k = \max(\tilde p_k, m_k)$ for all $k$.
  \State $c_k=\|\mathbf w_k\|_2^2$, \; $P_m=\sum_k c_k m_k^2$, \; $P_0=\sum_k c_k \hat p_k^2$.
  \If{$P_0 \le P_{\mathrm{tot}}$}
     \State \Return $\hat{\mathbf p}$.
  \Else
     \State $\alpha=\frac{P_{\mathrm{tot}}-P_m}{P_0-P_m}$;\;\; $\alpha\leftarrow \min\{1,\max\{0,\alpha\}\}$.
     \State $p_k=\sqrt{m_k^2+\alpha(\hat p_k^2-m_k^2)}$ for all $k$.
     \State \Return $\mathbf p$.
  \EndIf
\EndFunction
\For{{epoch $=1$ to max\_epochs}}
  \For{{each minibatch}}
    \State {$\tilde{\mathbf p}=\!\Call{Net}{\mathbf V,\{\gamma_k\},\{r_k\},P_{\mathrm{tot}}}$;\; $\mathbf p=\Call{Project}{\tilde{\mathbf p},\mathcal Q}$.}
    \State {$\ell = \ell_{\text{soft}}^{1}$ if $\mathcal Q{=}\mathcal K$ else $\ell_{\text{soft}}^{2}$;\; Adam update.}
  \EndFor
  \State {Validate EE;\; early stop if no improvement for $P$ epochs.}
\EndFor
\State {\textbf{Output:} trained weights $\Theta^\star$.}
\end{algorithmic}
\label{algorithm T}
\end{algorithm}

The proposed ANN method integrates inequality constraints into neural network training, enabling direct optimization from problem formulation equations without requiring labeled data, the framework of the ANN method is depicted in Fig.~\ref{fig:ANN}.
{The network minimizes the barrier-augmented EE objectives in (\ref{eq:lsoftp1})–(\ref{eq:lsoftp2}); feasibility is enforced by the in-loop projection operator described in Algorithm~\ref{algorithm T}.}
For the ANN method, we design dedicated loss functions corresponding to optimization problems $\mathcal{P}_1$ and $\mathcal{P}_2$ as
\begin{equation}
\begin{aligned}
\ell_{\text{soft}}^{1} &= {\text{minimize}}- \frac{B_w\sum_{k=1}^{K} \log_2 \left( 1 + \frac{\gamma_k |p_k|^2}{N_0} \right)}{P_{\text{com}}} -\\
&\left.\lambda \left( \sum_{k=1}^{K} \ln(p_k - p_{k_{\min}}+ \varepsilon)\right.  + \ln(P_{\text{tot}} - \sum_{k=1}^{K} \| p_k \mathbf{w}_k \|_2^2+\varepsilon) \right),
\end{aligned}
\label{eq:lsoftp1}
\end{equation}

\begin{equation}
\begin{aligned}
\ell_{\text{soft}}^{2} = {\text{minimize}}- &
\frac{B_w\sum_{k \in \mathcal{K} \setminus \mathcal{Q}} \log_2 \left( 1 + \frac{\gamma_k |p_k|^2}{N_0} \right)}{P_{\text{com}}}  \\
&
-\lambda \ln(P_{\text{tot}}' - \sum_{k \in \mathcal{K} \setminus \mathcal{Q}} \| p_k \mathbf{w}_k \|_2^2+\varepsilon),
\end{aligned}
\label{eq:lsoftp2}
\end{equation}
where $\lambda>0$ is the barrier weight, and $\varepsilon>0$ is a small constant (set to $10^{-6}$) to keep the log arguments strictly positive. 
In implementation, each slack term is evaluated as $\ln(\max\{x,\varepsilon\})$ to avoid numerical issues when the slack approaches zero.
{We adopt direct constrained learning with an in-loop feasibility projector (Algorithm~\ref{algorithm T}). At every forward pass we enforce $p_k\!\ge\!p_{k,\min}$ and $\|\mathbf W\mathbf p\|_2^2\!\le\!P_{\text{tot}}$, thereby satisfying (\ref{eq:27c}) and all per-user lower bounds. The network then minimizes the barrier-augmented EE objectives with penalty parameter $\lambda$. Training uses Adam with early stopping (patience $P{=}50$, max\_epochs $=2000$).}
The methodological novelty is twofold: 1) a lexicographic-style solver policy with a provably optimal greedy feasibility stage under ZF beamforming with statistical CSI (ZF+sCSI). 2) an in-the-loop capped-simplex feasibility projector that enforces all power and QoS constraints at every forward pass.

The input of neural network comprises the UPA response vector $\mathbf{V}$, HAP’s power budget $P_{\text{tot}}$, and QoS requirement $r_k$, while the output of the neural network is the optimized power coefficients $\{p_{k}\}$.
The parameters of the system utilized within the optimization procedure are delineated in Section V.
In the ANN method, a five-tier feedforward fully-connected neural network architecture is employed, characterized by sequential layer dimensions of $K$, 64, 64, 32, and 32 neurons.
{We ran a lightweight ablation over hidden-layer depth $\{2,3,4,5\}$ and width $\{32,64,128\}$.
After fine-tuning, a $4$-hidden-layer architecture with widths $[64,64,32,32]$ provides the best EE–stability trade-off and is adopted throughout.}
Each layer is followed by a rectified linear unit (ReLU) activation function except for the fifth layer.

The following Q3E algorithm presents the core procedures for solving the optimization problem $\mathcal{P}$.
Note that $\omega$ is only used in the weighted-sum baselines for comparison; Q3E follows a lexicographic-style solver policy rather than directly optimizing the weighted-sum objective.

\begin{algorithm}
\caption{QoS-Enhanced Energy-Efficient (Q3E) Beamforming Algorithm}
\begin{algorithmic}[1]
    \Require UPA response vector {$\{\mathbf{v}_k\}$}; HAP’s power budget $P_{\text{tot}}$; QoS requirement set $\{r_{k}\}$.
    \State Compute the beamforming vectors  {$\{\mathbf{w}_k\}$} based on the UPA response vector  {$\{\mathbf{v}_k\}$}.
    \State Calculate (\ref{eq:pmink}) to obtain $p_{k,\min}$.
    \If{Condition (\ref{eq:pemin}) is satisfied}
        \State Update $\mathcal{Q} = \mathcal{K}$ and solve problem $\mathcal{P}_1$ using either the ANN method or the SLSQP method to obtain $\{p_{k}\}$.
    \Else
        \State Calculate (\ref{calculate_Q}) to obtain $\mathcal{Q}$ and update $p_{k} = p_{k,\min}, \forall k \in \mathcal{Q}$.
        \State Solve problem $\mathcal{P}_2$ using either the ANN method or the SLSQP method to obtain $\{p_{k}\}$.
    \EndIf
    \Ensure The satisfied-user set $\mathcal{Q}$ and the optimized power coefficients $\{p_{k}\}$.
\end{algorithmic}
\label{algorithm P}
\end{algorithm}

Then, we evaluate the computational complexity of the algorithm. 
This complexity is primarily determined by whether all users meet their QoS requirements.
We assume that the probability of all users meeting their QoS requirements is $p$.
If all users achieve their QoS requirements, the complexity of solving problem $\mathcal{P}_1$ is related to the use of the ANN method and the SLSQP method.
The computational complexity of training neural network is $O\left( T_{1}\sum_{l=1}^{L} n_{l}^2 \right)$, where $T_1$ represents the number of training epochs, $L$ represents the number of network layers, and $n_l$ represents the number of neurons in the $l$-th layer.
The complexity of the SLSQP method is $O(T_{2}|\mathcal{K}|^3)$, where $T_2$ is the number of SLSQP iterations.
If some users cannot achieve their QoS constraints, the complexity of solving problem $\mathcal{P}_2$ is related to the ANN method and the SLSQP method, as well as the number of users not meeting QoS requirements.
The SLSQP method is similar to the situation where all users achieve their QoS requirements, but it only optimizes for some users, so its complexity is $O(T_{2}(|\mathcal{K}|-|\mathcal{Q}|)^3)$. 
In addition, the neural network training is the same as the situation where all users achieve their QoS requirements.
Besides, a quick sort approach is utilized to obtain (\ref{ascending_order_pk}), the computational complexity is then $O(|\mathcal{K}|\log |\mathcal{K}|)$.
Overall, the computational complexity of Algorithm 1 is $O( N_{t}^3 + |\mathcal{K}| + p\min\{(T_{1}\sum_{l=1}^{L} n_{l}^2,T_{2}|\mathcal{K}|^3 \}+(1-p)|\mathcal{K}| \log |\mathcal{K}|\min\{T_{1}\sum_{l=1}^{L} n_{l}^2,T_{2}(|\mathcal{K}|-|\mathcal{Q}|)^3 )\}$.

\section{Numerical Simulation Results}
{In this section, we assess the accuracy of the proposed propulsion power-consumption model and evaluate the performance of the proposed beamforming algorithms for the HAP communication system.}

\subsection{Comparison Algorithms and Parameter Setting}
To verify the accuracy of the proposed model, we compare it with the model in \cite{SONG2024109266} and with CFD numerical analysis results obtained using STAR-CCM+ software.
{All CFD validations use incompressible steady RANS with SST $k$–$\omega$ at ISA-1976 properties for $h=20$\,km (Mach $<0.1$).} 
As previously stated, the proposed algorithm prioritizes user QoS satisfaction ratio and EE.
{To validate its performance, we evaluate four benchmark algorithms: 
1) a QoS-satisfaction beamforming algorithm that maximizes the number of users meeting their QoS targets under a power constraint \cite{10636212}. 
2) a max-$R_k$ algorithm that maximizes the sum rate under the same constraint \cite{10636212}. 
3) a Proximal Policy Optimization (PPO) benchmark that learns to optimize the composite objective under the power constraint \cite{11162243}. 
4) an Ant Colony Optimization for Continuous Domains (ACOR) benchmark that searches the continuous power-allocation space to optimize the same objective \cite{THAT2026129401}.}
The comparison algorithms are implemented on PyTorch 2.5, SciPy 1.9 and Python 3.9, and a PC configured with Core i7, 16G RAM, and Windows 10 OS is used.

We consider a HAP operating at 20\,km altitude \cite{10445467}, stationed at a predefined location to provide communication services for ground users with three distinct QoS levels \cite{10474118}: high, medium, and low.
The QoS thresholds for these levels are defined as:
High QoS: 60\,Mbps $<$ Data rate $\leq$ 90\,Mbps, 
Medium QoS: 30\,Mbps $<$ Data rate $\leq$ 60\,Mbps, 
Low QoS: Data rate $\leq$ 30\,Mbps. 
The parameters related to the HAP are set as follows: the length $l= 140$\,m, the width $d =34$\,m, the total volume $\Omega = 85000$ m$^3$, the tail drag correction factor $K_F = 1.12$, the efficiency of motor ${\eta_m = 0.85}$, and the maximum flight airspeed is 25\,m/s.
The environmental parameters follow ISA-1976 \cite{QI2025620} and are set as follows: the air density $\rho=0.08803$\,kg/{m}${^{3}}$, 
and the {dynamic viscosity} $\mu = 1.4216 \times {10}^{\text{-5}}$\,Pa$\cdot$s.
In addition, each propeller has three blades(3\,m radius for each blade), and the distance between a propeller and the HAP hull is 0.72\,m.
The parameters related to HAP communications are set as follows \cite{9852292,10445467}: system bandwidth $B_w = 10$\,MHz, Rician factor $\kappa_k = 12$\,dB, number of antennas $N^x \times N^y = 12 \times 12$, number of users $K = 9$, antenna spacing $r^x = r^y = \lambda/2$ with $\lambda = c/f_c$, inefficiency of power amplifier $ \xi = 2$, static power consumption of baseband beamformer $ P_{\text{BB}} = 200$\,mW, static power consumption of local oscillator $ P_{\text{LO}} = 5$\,mW, per-RF chain power consumption $ P_{\text{RFC}} = 338$\,mW. Besides, the channel power is defined as $\gamma_{k} = G_{\text{HAP}} G_{\text{user}} N_{t} \left(\frac{c}{4\pi f_{c} h}\right)^{2}$ with the antenna gains of the transmitter and the receiver $G_{\text{HAP}} = G_{\text{user}}= 3$\,dB, carrier frequency $f_{c} = 2100$\,MHz, and the speed of light $c = 3 \times 10^8$\,m/s.

\subsection{Performance Evaluation}
\subsubsection{Propulsion Power Consumption Model Simulation}
\begin{figure}[!t]
	\centering
	\includegraphics[width=0.8\linewidth]{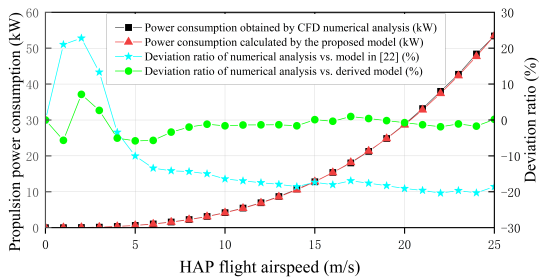}
	\caption{{Propulsion power and deviation ratios for the derived model.}}
	\label{fig:haps-required-energy}
\end{figure}

We simulate the task scenario of a HAP performing wind-resistant flight for station-keeping above a designated location. 
We plot the tendency of the propulsion power consumption over the HAP flight airspeed and the deviation ratios of different propulsion power consumption models in Fig.~\ref{fig:haps-required-energy}. 
From this figure, we can observe: 
1) Propulsion power consumption increases significantly with flight airspeed, exhibiting a superlinear growth pattern. 
2) The analytical results obtained from the derived model show close agreement with numerical simulations, demonstrating better accuracy than existing models. 
3) The maximum deviation ratio of the derived model remains below 6\%, with deviations constrained within 5\% when airspeed exceeds 3 m/s. 
This represents a substantial improvement over the model in \cite{SONG2024109266} which shows deviations exceeding 20\%. 
4) Fig.~\ref{fig:propulsion energy consumption model deviation} reveals an additional energy penalty caused by aerodynamic interference between the HAP hull and propulsion system, particularly noticeable at higher airspeeds.

To further evaluate the model accuracy, a comparative analysis was conducted between the derived model, the benchmark model in \cite{SONG2024109266}, and the CFD numerical analysis results.
The derived model shows the closest agreement with the CFD results.
In particular, it achieves an average 84.3\% reduction in propulsion power deviation compared with the benchmark model.

\subsubsection{Beamforming Design Algorithms Simulation}
\begin{figure}[!t]
	\centering
	\includegraphics[width=0.8\linewidth]{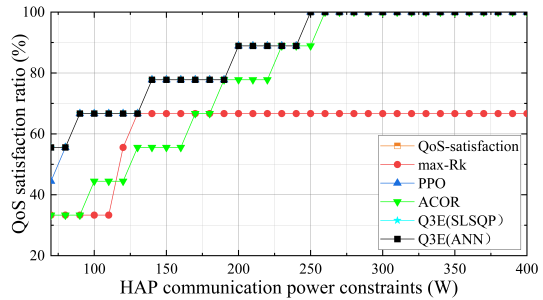}
	\caption{{The QoS satisfaction ratio versus communication power constraints for {five} beamforming methods.}}
	\label{fig:QoS rate}
\end{figure}

\begin{figure}[!t]
	\centering
	\includegraphics[width=0.8\linewidth]{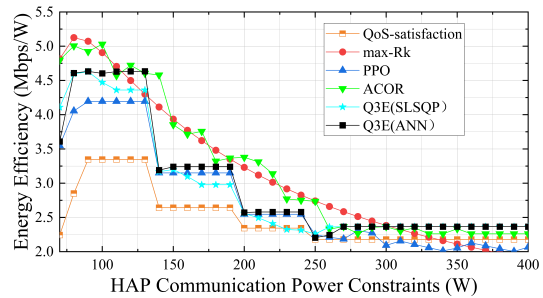}
	\caption{{The EE performance versus communication power constraints for {five} beamforming methods.}}
	\label{fig:HAP EE 3}
\end{figure}

With an accurate propulsion power consumption model, we can obtain an accurate communication power constraint. 
Next, a realistic scenario is simulated, where a HAP provides communication services to ground users with three distinct QoS requirement levels, each corresponding to three user terminals.
{As shown in Figs.~\ref{fig:QoS rate} and~\ref{fig:HAP EE 3}, the Q3E algorithm is evaluated against four benchmarks: QoS-satisfaction, max-$R_k$, PPO, and ACOR.}
{The ablation in Table~\ref{tab:ablation} shows that removing the scaling module yields frequent infeasibility under tight budgets (hence EE is not reported), while removing the soft-loss module preserves feasibility but slightly reduces EE relative to full Q3E(ANN).}

\begin{table}[t]
\centering
\caption{Ablation on soft-loss and scaling.}
\label{tab:ablation}
\small
\setlength{\tabcolsep}{4pt}
\resizebox{\columnwidth}{!}{%
\begin{tabular}{lccc}
\toprule
Variant & Feasibility (\%) & Mean-Overshoot (W) & Mean EE (Mbps/W) \\
\midrule
Q3E (ANN) & \textbf{100.0} & 0.00 & 2.976 \\
Ablation-1 (no soft loss) & \textbf{100.0} & 0.00 & 2.822 \\
Ablation-2 (no scale) & 64.7 & 37.9 & --- \\
\bottomrule
\end{tabular}%
}
\end{table}

Fig.~\ref{fig:QoS rate} shows that Q3E, QoS-satisfaction, and PPO exhibit similar monotonic gains with increasing power and reach full satisfaction around 250~W, while ACOR rises more gradually and catches up as the budget relaxes. By contrast, max-$R_k$ plateaus near 66\% over 70–400~W since it optimizes sum-rate rather than the count of QoS-satisfied users.

Fig.~\ref{fig:HAP EE 3} compares EE from 70 to 600~W. Key findings:
1) \textbf{Q3E versus QoS-satisfaction:} Q3E is about 35–45\% higher at 90–150~W and about 8–10\% higher at 300~W or above because it optimizes EE rather than only the satisfied-user count.
2) \textbf{Q3E versus PPO:} Both show steps near 140, 200, and 250~W as more QoS users activate. 
Q3E is higher by about 5–15\% under tight to mid constraints and about 10–25\% when the budget is 300~W or above.
3) \textbf{Q3E versus ACOR:} Around 150~W, Q3E is about 15\% lower. 
For budgets of 300~W or above, they converge, with Q3E higher on average by 1–4\%.
4) \textbf{Q3E versus max-$R_k$:} Q3E is 8–12\% lower at 70–110~W and up to about 15\% lower at 140–300~W. For budgets of 300~W or above, Q3E is 10–25\% higher, consistent with max-$R_k$ continuing to chase throughput when it is no longer EE-optimal.
5) \textbf{Q3E(ANN) versus Q3E(SLSQP):} Both variants show the same stepwise trend as additional QoS users become active. Q3E(SLSQP) is slightly better only at 70, 250, and 260~W. At all other budgets it falls below Q3E(ANN).

Overall, the interactive generative AI scheme facilitates solving interdisciplinary problems by seamlessly integrating domain-specific knowledge.
The neural network has a built-in feasibility projector. 
It also uses a barrier loss. 
These features make sure the solution always meets the constraints during training and deployment.

\section{Conclusion}

This paper studied power allocation for HAP communications by jointly modeling propulsion power and optimizing communication power. 
Guided by a generative AI agent, we developed an accurate propulsion power model that captures hull-propeller aerodynamic interference via aerodynamic principles and numerical analysis. 
Compared with the benchmark, it reduces the average deviation by 84.3\%; at $V_0{=}25$~m/s, the error drops from 8367.3~W to 85.32~W (98{:}1), enabling substantially tighter system power budgeting and more reliable communication power allocation.
Building on this model, we formulated a beamforming optimization problem and proposed an ANN-based constrained training approach that requires no supervised datasets. 
Under the same communication-power constraints, Q3E achieves the highest QoS satisfaction ratio among the evaluated baselines, and typically attains higher energy efficiency (EE) at the same QoS satisfaction ratio. 
Overall, simulations validate both the propulsion-power modeling accuracy and the effectiveness of the proposed beamforming optimization.

This study targets a single finalized HAP and a static sCSI model; multi-HAP cooperation and heterogeneous use cases are out of scope. 
Next, we will extend to mobile/heterogeneous scenarios with cooperative multi-HAP beamforming and co-optimize power and trajectory for joint aero–propulsion–communication design.

\section{Acknowledgment}
Any opinions, findings and conclusions or recommendations expressed in this material are those of the author(s) and do not reflect the views of National Research Foundation, Singapore.

\bibliography{IEEEabrv,Power_Allocation_for_HAP}
\bibliographystyle{IEEEtran}
\end{document}